\documentclass[10pt]{article}
\usepackage[english]{babel}
\usepackage{amsmath}
\usepackage{amsfonts}
\usepackage{amssymb}
\usepackage{dsfont}
\usepackage{color} 
\usepackage[svgnames,dvipsnames]{xcolor}
\usepackage{amsthm}
\usepackage{authblk}
\usepackage[left=2cm,right=2cm,top=3cm,bottom=3cm]{geometry}
\usepackage{graphicx}
\usepackage{tikz}
\usetikzlibrary{shapes,arrows,calc}
\usetikzlibrary{shapes.geometric}
\usetikzlibrary{positioning}
\usepackage{pgfplots}
\usepackage{pstricks}
\usepackage{listings}
\usepackage{subcaption}
\usepackage{booktabs}
\usepackage{pifont}
\usepackage{setspace}
\usepackage[hidelinks]{hyperref}
\usepackage{csquotes}

\tikzstyle{block} = [rectangle, draw, fill=green!20,
text width=6em, text centered, minimum height=2em,node distance= 3cm]
\tikzstyle{block2} = [rectangle, draw, fill=RoyalBlue!20,
text width=10em, text centered, minimum height=4em,node distance= 3cm]
\tikzstyle{block3} = [rectangle, draw, fill=RoyalBlue!20,
text width=12em, text centered, minimum height=4em,node distance= 3cm]
\tikzstyle{line} = [draw, -latex']
\tikzstyle{cloud} = [draw, rectangle,fill=white, node distance=3cm,
minimum height=2em]
\tikzstyle{cercle} = [draw, circle, fill= white,node distance = 3cm, minimum height = 2em]
\tikzstyle{cercle2} = [draw,circle,fill=green!10,node distance = 2cm, text width = 5em, text centered, minimum height = 2em]
\tikzstyle{fichier} = [draw,rectangle,fill=pink!10,text centered,text width=100em,minimum height=4em,node distance=3cm]
\tikzstyle{reseau}=[draw,rectangle,fill=RoyalBlue!20,text width=10em, text centered,minimum height=4em,node distance=4em]
\tikzstyle{VertexStyle} = [shape            = ellipse,
minimum width    = 6ex,%
draw]
\tikzstyle{EdgeStyle}   = [->,>=stealth']

\newcommand{\pte}{\operatorname{PTE}}
\newcommand{\nde}{\operatorname{NDE}}
\newcommand{\nie}{\operatorname{NIE}}
\newcommand{\tte}{\operatorname{TTE}}

\usepackage[natbib=true,bibencoding=utf8,hyperref=true,backref=false,
backrefstyle=three,bibstyle=authoryear,citestyle=authoryear,url=false,
maxcitenames=2,maxbibnames=3,dashed=false,isbn=false,doi=false,
giveninits=true,uniquename=init]{biblatex} 
\DeclareNameAlias{default}{family-given}
\DeclareFieldFormat[article]{title}{#1}
\DeclareFieldFormat[book]{title}{\textit{#1}} 
\DeclareFieldFormat[incollection]{title}{#1} 
\DeclareFieldFormat[inbook]{title}{#1} 
\DeclareFieldFormat[collection]{title}{#1} 
\DeclareFieldFormat[book]{publisher}{\textit{#1}}
\DeclareFieldFormat[online]{url}{\small \url{#1}}
\DeclareFieldFormat{pages}{#1}
\DeclareFieldFormat[article]{volume}{\textbf{#1}}

\renewbibmacro*{volume+number+eid}{%
	\printfield{volume}%
	\setunit*{\addnbspace}
	\printfield{number}%
	\setunit{\addcomma\space}%
	\printfield{eid}}
\DeclareFieldFormat[article]{number}{\mkbibparens{#1}}

\AtEveryBibitem{%
	\clearfield{urldate}%
	\clearfield{ISSN}%
	\clearfield{abstract}%
	\clearfield{number}%
}
\DeclareFieldFormat{citehyperref}{%
	\DeclareFieldAlias{bibhyperref}{noformat}%
	\bibhyperref{#1}}
\DeclareFieldFormat{textcitehyperref}{%
	\DeclareFieldAlias{bibhyperref}{noformat}%
	\bibhyperref{%
		#1%
		\ifbool{cbx:parens}
		{\bibcloseparen\global\boolfalse{cbx:parens}}
		{}}}
\savebibmacro{cite}
\savebibmacro{textcite}
\renewbibmacro*{cite}{%
	\printtext[citehyperref]{%
		\restorebibmacro{cite}%
		\usebibmacro{cite}}}
\renewbibmacro*{textcite}{%
	\ifboolexpr{
		( not test {\iffieldundef{prenote}} and
		test {\ifnumequal{\value{citecount}}{1}} )
		or
		( not test {\iffieldundef{postnote}} and
		test {\ifnumequal{\value{citecount}}{\value{citetotal}}} )
	}
	{\DeclareFieldAlias{textcitehyperref}{noformat}}
	{}%
	\printtext[textcitehyperref]{%
		\restorebibmacro{textcite}%
		\usebibmacro{textcite}}}

\renewbibmacro{in:}{}

\addto\captionsenglish{%
}
\usepackage{color}
\usepackage{fancyvrb}

\DefineVerbatimEnvironment{Highlighting}{Verbatim}{commandchars=\\\{\}}
\usepackage{framed}
\definecolor{shadecolor}{RGB}{255,255,255}

\pgfplotsset{compat=1.18}
\begin{document}

\title{Tutorial for Surrogate Endpoint Validation Using Joint modeling and Mediation Analysis}
\author[1]{Quentin Le Coent}
\author[2]{Catherine Legrand}
\author[3]{Virginie Rondeau}
\affil[1]{Bloomberg School of Public Healt, Johns Hopkins University}
\affil[2]{ISBA/LIDAM, UCLouvain}
\affil[3]{Bordeaux Population Health Research Center, INSERM U1219, Université de Bordeaux}
\date{}

\maketitle 

\abstract{
The use of valid surrogate endpoints is an important stake in clinical research to help reduce both the duration and cost of a clinical trial and speed up the evaluation of interesting treatments. Several methods have been proposed in the statistical literature to validate putative surrogate endpoints. Two main approaches have been proposed: the meta-analytic approach and the mediation analysis approach. The former uses data from meta-analyses to derive associations measures between the surrogate and the final endpoint at the individual and trial levels. The latter rather uses the proportion of the treatment effect on the final endpoint through the surrogate as a measure of surrogacy in a causal inference framework. Both approaches have remained separated as the meta-analytic approach does not estimate the treatment effect on the final endpoint through the surrogate while the mediation analysis approach have been limited to single-trial setting. However, these two approaches are complementary. In this work we propose an approach that combines the meta-analytic and mediation analysis approaches using joint modeling for surrogate validation. We focus on the cases where the final endpoint is a time-to-event endpoint (such as time-to-death) and the surrogate is either a time-to-event or a longitudinal biomarker. Two new joint models were proposed depending on the nature of the surrogate. These model are implemented in the \verb|R| package \verb|frailtypack|. We illustrate the developed approaches in three applications on real datasets in oncology. 
}

\section{Introduction}

Given the continued development of new and more effective therapies in
oncology, the use of standard (final) endpoints in clinical trials may
affect the feasibility of these trials. For example, a significant
increase in overall survival requires long-term follow-up of patients to
ensure adequate statistical power and therefore increases the duration
and cost of the trial. The use of surrogate endpoints is therefore a
possible way to enable such otherwise unfeasible studies. Surrogate
endpoints are intermediate endpoints measured earlier and/or more
frequently than the final (true) endpoints that help reduce the duration
and cost of a trial compared to final endpoints
(\citet{buyse2000validation}). The benefit of a surrogate endpoint is
that is can be used as a primary endpoint in place of the final endpoint
in a new clinical trial. However, the final endpoint being the
``gold-standard'', conclusions of treatment efficacy in a clinical trial
based solely on its observed efficacy on the surrogate should be
extendable to the final endpoint. Therefore, prior to any use in a new
trial, a surrogate endpoint must have been carefully validated. This
validation requires that both the final endpoint and the surrogate have
been studied in previous studies. Several methods have been proposed in
the last decades following the seminal paper of Prentice
\citep{prentice1989surrogate}. Since, two main approaches have been
developed, one based on quantifying ``associations'' between the
surrogate and the final endpoint and others based on identification of
the proportion of treatment effects on the final endpoint through the
surrogate \citep{joffe2009related}. The main developments of the
``associations'' approach is the meta-analytic approach
\citep{daniels1997meta,buyse2000validation} while the proportion of
treatment effect has been developed in the mediation analysis setting
\citep{vandenberghe2018surrogate,weir2022counterfactual,parast2017evaluating}.

The meta-analytic approach uses data from meta-analyses of randomized
clinical trials to derive and quantify the association between the
surrogate and the final endpoint at two levels. At the individual level
the association between the two endpoints is quantified while at the
trial-level the association between the treatment effects on both
endpoints across the trials is used. For the individual-level
association, this approach relies on joint models for the two endpoints
using individual (patient) level random effects. These random effects
can be shared or correlated between the two endpoints and can be used to
estimate an individual-level association between the surrogate and the
final endpoint. For the trial-level association, methods based on one or
two stages approaches have been developed. In the two-stage approaches,
the joint model is estimated with fixed trial-specific treatment effects
on the two endpoints. Then at the second stage the estimated treatment
effects on the two endpoints for each trial can be regressed in a linear
mixed model. The resulting coefficient of determination can then be used
as trial-level measure of association between the treatment effects.
However, these two stage approaches can be replaced by one stage
approaches in which the fixed trial-specific treatment effects in the
joint model are replaced by the sum of a fixed effect (not
trial-specific) and a trial-level random effect for each endpoint. Then
the elements of the covariance matrix of these trial-level random
effects are used to derive the coefficient of determination between the
treatment effects. Then, a surrogate endpoint will be claimed validated
if the associations at both individual and trial levels are sufficiently
large. The meta-analytic approach has been broadly studied and adapted
to settings with different types of surrogate and final endpoint:
continuous, binary, censored time-to-event
\citep{burzykowski2005evaluation}.
Another approach, less developed, focuses on identifying the pathways
through which the treatment effect on the final endpoint goes. A natural
surrogacy measure in that case is the proportion of treatment effect,
defined as the ratio of the indirect treatment effect on the final
endpoint through the surrogate over the total treatment effect on the
final endpoint. Early developments of this approach were based on
estimating regression coefficients of the treatment effect in a model
for the final endpoint adjusted only on the treatment arm and a model
adjusting on both the treatment and the surrogate
\citep{prentice1989surrogate,freedman1992statistical}. Then the ratio of
the estimated regression coefficient associated with the treatment in
the second model over the coefficient estimated in the first model was
taken as a measure of the proportion of treatment effect. However, this
approach lacks causal interpretability since adjusting on a
posttreatment covariate affected by the treatment (the surrogate) may
yield biased estimation of the treatment effect by neglecting potential
confounding between the surrogate and the final endpoint
\citep{rosenbaum1984consequences,robins1986new,frangakis2002principal}.
Moreover, using estimated regression coefficients to define a proportion
of treatment effect complicates the interpretation of this proportion on
different scales. For example, a high proportion using estimated
coefficients in a Cox model might not translates in the same proportion
on the survival scale (i.e.~by using the difference of survival rates as
a measure of the treatment effect). Additionally, estimates of hazard
models may also suffer from a lack of causal interpretability in
contrary to survival measures \citep{hernan2010hazards}. Therefore,
methods based on causal mediation analysis have been developed. These
methods often rely on the causal framework of potential outcomes
\citep{holland1986statistics}. The benefit of using causal mediation
analysis is two fold: first it does not suffer from a lack of causal
interpretability; second, it gives model-free definitions of ``treatment
effects'', which can be defined in a broad variety of situations. For
example, if we are interested in the case where the final endpoint is a
time-to-event, then it is possible to give a definition of the
proportion of treatment effects on the survival scale. However, causal
mediation analysis comes at a price: one often needs to rely on strong
identifiability assumptions \citep{imai2010identification} and the
estimations of the quantities of interest are generally more complex
than simply fitting a regression model. Less developments of the
mediation analysis for surrogate endpoints validation have been made
compared to the meta-analytic approach. One limitation of the mediation
analysis is that it is often limited to single-trial data, which may
complicate the generalization of its results.

We propose a novel development that combines the meta-analytic and
mediation analysis approaches. We focus on the specific case where the
final endpoint is a time-to-event. We develop two joint models for the
case where the surrogate is also a time-to-event or a continuous
longitudinal biomarker. The first model is an extension of the joint
model proposed by \citet{sofeu2019one} that allows for a potential
mediated treatment effect on the final endpoint through the surrogate.
The second model is a joint model for a time-to-event and a longitudinal
surrogate. In both models the use of individual and trial-level random
effects allows for i) using meta-analytic data and ii) deriving
individual and trial-level associations. From these models we also gave
a causal definition of the indirect treatment effect on the final
endpoint through the surrogate and the direct treatment effect using
mediation analysis. The use of random effects allows us to make less
stringent identifiability assumptions than often found in the mediation
analysis literature.

This paper is organized as follows. In the first Section we provide a
review on the existing \verb|R| packages available on the
\verb|CRAN| for surrogate endpoints validation with their principal
options (type of endpoints, methodology etc.). In a second section we
introduce the statistical methodology for surrogate endpoint validation.
Two joint models are presented, the first one for the case where the
surrogate endpoint is a time-to-event and the second where the surrogate
is a longitudinal continuous biomarker. In both models the final
endpoint will be a time-to-event. We also present the measures of
surrogacy resulting from these models. In the third Section we present
two functions implemented in \verb|frailtypack|,
\verb|jointSurroPenal| and \verb|longiPenal|, dedicated to each
proposed model. We present the main arguments of these functions and
their role. In the fourth section we provide three illustrations of the
proposed models on real-data examples in oncology. Finally in the fifth
section we conclude and provide some further developments.

\section{Existing R packages for surrogate validation}

In this Section we review some \verb|R| packages which provide useful
measures for surrogate endpoint validation.

The package \verb|Rsurrogate| provides several functions to compute
the proportion of treatment effect on a final endpoint through a
surrogate. It can be used with a censored time-to-event final endpoint
and a surrogate measured at a specific time, with the surrogate being
treated as missing if the final endpoint occurs before this time point.
It can also be used to compute the proportion of treatment effect on a
(time-to-event) final endpoint through several surrogates. It further
provides a function to compute this proportion in the case where the
final endpoint is a continuous endpoint, where a function to correct for
a potential measurement error in the surrogate is also available. In all
cases the surrogate (or vector of surrogates) is assumed to be
continuous \citep{rsurrogate}.

The package \verb|SurrogateOutcome| provides a function to estimate
the proportion of treatment effect on a time-to-event endpoint through a
time-to-event surrogate observed up to a landmark time
\citep{SurrogateOutcome}. These effects are defined as differences in
restricted mean survival times. This package also provides a function in
which the surrogate is replaced by the final endpoint itself observed up
to this landmark time. The package \verb|longsurr| provide a function
to compute the proportion of treatment effect on a final continuous
endpoint through a longitudinal surrogate, where it is assumed that each
patient has the same number of repeated observations of the surrogate
(and that each measurement is taken at the same time) \citep{longsurr}.
The package \verb|surrogate| \citep{surrogate} is the most developed
package dedicated to surrogate endpoint validation available on the
\verb|CRAN|. It provides numerous functions to assess surrogacy
depending on the nature of the surrogate and the final endpoint based on
the meta-analytic, information-theoretic, and causal-inference
frameworks. For example, this package can be used to assess the
trial-level surrogacy for two time-to-event endpoints using a two-stage
approach or based on information theory and a two-stage approach. It can
also be used to estimate the trial-level surrogacy by providing directly
the coefficients associated with the estimated treatment effects on the
final and surrogate endpoints. It also provides a function to evaluate
surrogacy following Prentice's criteria in a single-trial setting with
two continuous endpoints. Although we succinctly presented packages
developed in \verb|R|, implementations for surrogate validation have
also been proposed in other softwares such as \verb|SAS|
\citep{alonso2016applied}. None of those packages combined the one-step
meta-analytic approach with the mediation analysis. We propose an
\verb|R| package to validate a longitudinal biomarker or failure time
as a surrogate of a time-to-event using mediation analysis and the
meta-analytic approach.

 \section{Methods: surrogate validation using joint modeling and mediation analysis}

In this Section we introduce the two main models: the first one for a
time-to-event surrogate and the second one for a longitudinal surrogate.
The surrogacy evaluation criteria are presented for both the
meta-analytic and the mediation approaches. Estimation method of model
parameters is also presented.

\subsection{Models and estimation}

In a setting of meta-analytic data with \(K\) trials, let \(i\) denotes
the \(i\)th trial and \(ij\) the \(j\)th subject from trial \(i\),
\(j=1,\dots,n_i\). The treatment and covariates are denoted by
\(Z_{ij}\) and \(X_{ij}\).

\subsubsection{Time-to-event surrogate}

Let \(T_{ij}\) and \(S_{ij}\) be the final endpoint and the surrogate
endpoint respectively. Let \(C_{ij}\) denotes the censoring time. We
assume a semi-competing risks setting in that \(S_{ij}\) can be censored
by \(T_{ij}\) but not the opposite. The observed outcomes are therefore
\(T^*_{ij} = \min(T_{ij},S_{ij})\),
\(\delta_{ij} = \mathds{1}_{T_{ij} \leq C_{ij}}\),
\(S^*_{ij} = \min(S_{ij},T^*_{ij})\) and
\(d_{ij} = \mathds{1}_{S_{ij} \leq T^*_{ij}}\). The joint model is given
by \citep{lecoent2022time}: \begin{equation}\label{eq:model1}
\left\{
\begin{array}{l}
  \lambda_{S}\left(t \mid \omega_{ij},u_i,\nu_{S,i},Z_{ij},X_{ij} \right) = \lambda_{0,S}(t) \exp \Big{(} \omega_{ij} +   u_i + Z_{ij}(\nu_{S,i} + \beta_{Z,S}) + \beta_S' X_{ij}  \Big{)} \\[0.2cm]
  \lambda_T\left (t \mid \omega_{ij},u_i,\nu_{T,i},Z_{ij},S_{ij},X_{ij} \right ) = \lambda_{0,T}(t) \exp \Big{(}  \zeta \omega_{ij} +  \alpha u_i + Z_{ij}(\nu_{T,i}+ \beta_{Z,T})  \\ \hspace{20.84em}  +\beta_T'X_{ij} + \gamma(S_{ij}) I\left(S_{ij} \leq t \right) \Big{)}.
\end{array}
\right.\end{equation}\\
In this model, \(\omega_{ij}\) and \(u_i\) are individual and trial
level random effects respectively used to take into account the
heterogeneity between patients at the individual and trial levels. The
trial-level random effects \((\nu_{S,i}, \nu_{T,i})\) take into account
a potential heterogeneity between treatment effects across trials. All
of these random effects are assumed to be Gaussian: \begin{equation*}
\begin{split}
&\omega_{ij} \sim \mathcal N(0, \theta^2) \\
&u_i \sim \mathcal N(0, \gamma^2) \\
& (\nu_{S,i},\nu_{T,i})' \sim \mathcal N(0, \Sigma_{\nu}).
\end{split}
\end{equation*}

The parameters \((\zeta, \alpha)\) are power parameters allowing the
random effects \(\omega_{ij}\) and \(u_i\) to affect the hazard
functions of \(S\) and \(T\) differently. Parameters \(\beta_{Z,S}\) and
\(\beta_{Z,T}\) are the fixed treatment effects on the surrogate and the
final endpoint respectively. The functions \(\lambda_{0,T}\) and
\(\lambda_{0,S}\) are the baseline hazard functions for the two
endpoints.

In model (\ref{eq:model1}), the term
\(\gamma(S_{ij}) I\left(S_{ij} \leq t \right)\) corresponds to a direct
link of the surrogate on the final endpoint. This link can be
interpreted as a modification of the hazard function of \(T_{ij}\) at
the time of occurrence of \(S_{ij}\) whose magnitude is given by the
function \(\gamma(\cdot)\). From a mediation perspective this term is
important as without it the two endpoints would be independent given the
random effects, treatment and covariates, hence no direct effect of the
surrogate on the final endpoint would be possible (which corresponds to
removing the arrow between \(S\) and \(T\) in Figure \ref{fig:fig1}(b)).

Outside this mediation perspective, and based only on the meta-analytic
approach, a direct link between \(S\) and \(T\) is not required, and a
joint model without the term
\(\gamma(S_{ij}) I\left(S_{ij} \leq t \right)\) can be used
\citep{sofeu2019one}.

\subsubsection{Longitudinal surrogate}

Let \(M_{ij}\) denotes the surrogate which is a continuous longitudinal
biomarker (repeated measures over time) and \(T_{ij}\) is a
time-to-event defined as previously. For subject \(ij\) we observe
\(\tilde M_{ij} = \{ \tilde M_{ij}(t_{ijk}),1\leq k \leq n_{ij}\}\) the
longitudinal biomarker measured with an error. where \(n_i{ij}\) is the
number of repeated measurements of subject \(ij\).

The proposed joint model is given by, \begin{equation}\label{eq:model2}
\left\{
\begin{array}{l}
  \tilde{M}_{ij}(t) = 
  \theta_{ij}^{'} f(t) + \beta_M' X^M_{ij}(t)  + (\beta_{Z,M}+\nu_{M,i}) Z_{ij} + \varepsilon_{ij}(t) \\[0.2cm]
  \lambda_{ij}(t) = \lambda_{0}(t) \exp\left( \left (\beta_{Z,T} + \nu_{T,i} \right ) Z_{ij} + \beta_T' X^T_{ij}(t) + \eta' h \left (M_{ij} \right ) \right)
\end{array}
\right.\end{equation}

where
\(M_{ij}(t) = \theta_{ij}' f(t) + \beta_M' X^M_{ij}(t) + (\beta_{Z,M}+\nu_{M,i}) z\)
is the true, error-free, marker and
\(\varepsilon_{ij}(t) \sim \mathcal{N}(0, \sigma_{\varepsilon}^2)\) (the
independent measurement error).

As in Model (\ref{eq:model1}), the random effects
\((\nu_{S,i},\nu_{T,i})\) are trial-level effects taking into account
the heterogeneity of the treatment effect across trial and are assumed
to be jointly Gaussian:
\((\nu_{M,i},\nu_{T,i})' \sim \mathcal{N} \left( 0, \Sigma_\nu \right)\).
The function \(f(t)\) is a vector that represents the temporal evolution
of the biomarker and may be composed of several components. The vector
\(\theta\) is the sum of fixed effects and individual random-effects
associated with each component of \(f(t)\),
\(\theta_{ij} = \beta + \omega_{ij}\). For example if
\(f(t) = \left(1,t \right)'\) then
\(\theta_{ij} = \left (\beta_0 + \omega_{ij0} , \beta_1 + \omega_{ij1} \right)'\).
The vector \(\omega_{ij}\) is assumed to be Gaussian with mean \(0\) and
unstructured covariance matrix \(D\), \(\omega \sim \mathcal N(0,D)\).

In Model (\ref{eq:model2}), the two submodels are linked together
through the ``link'' function \(h \left ( M_{ij} \right )\). This
function represents the association structure between the longitudinal
surrogate and the final endpoint. The strength of this association is
taken into account through the ``association parameter'' \(\eta\). This
parameter quantifies the magnitude of the dependence between both
outcomes. The most common link functions found in the literature are the
following \citep{rizopoulos2012joint}:

\begin{itemize}
    \item[$\bullet$] The ``current level'' link: $h \left ( M_{ij} \right ) = M_{ij}(t)$. This is probably the most natural link between the biomarker and the time-to-event that assumes that the risk of occurrence of $T$ at any time $t$ depends only on the current value of the  biomarker at that time. 
    \item[$\bullet$] The ``current slope'' link: $h \left (M_{ij} \right ) = \frac{\mathrm{d}M_{ij}(t)}{\mathrm{d}t}$. Rather than considering an association between the hazard function of $T$ with the current value of $M_{ij}$ at time $t$, we consider an association with the increase (or decrease) of $M_{ij}$ at time $t$. 
    \item[$\bullet$] $h \left ( M_{ij} \right ) = \omega_{ij}$, the ``shared random-effects'' link function. As it is time-independent, it assumes that dependence structure between the two outcomes is constant over time regardless of the biomarker evolution.
\end{itemize}

\begin{figure}
\begin{subfigure}[b]{\textwidth}
    \centering
    \includegraphics[width=8cm]{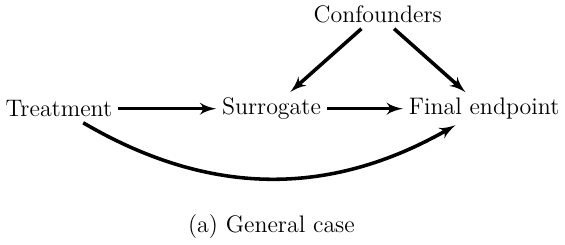}
\end{subfigure}
\hfill
\begin{subfigure}[b]{\textwidth}
    \centering
    \includegraphics[width=8cm]{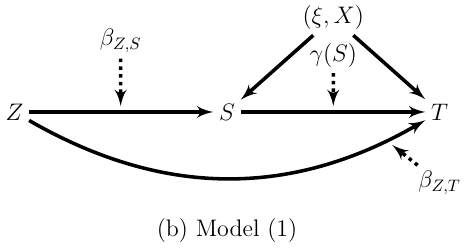}
\end{subfigure}
\hfill 
\begin{subfigure}[b]{\textwidth}
    \centering
    \includegraphics[width=8 cm]{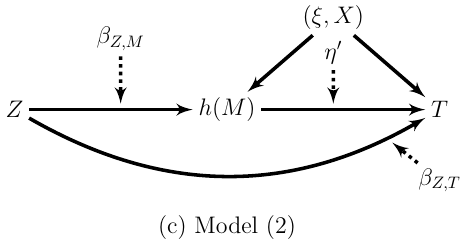}
\end{subfigure}
\caption{Directed acyclic graph associated with the joints models: (a) in a general case, (b) for a time-to-event surrogate as represented by Model (1) and (c) for a longitudinal surrogate (Model (2)). Here $\xi$ denotes the vector of all the random effects in models 1 and 2. \label{fig:fig1}}
\end{figure}

From a mediation viewpoint, in order to enable a potential mediated
treatment effect on the final endpoint through the surrogate, the
quantity in the link function must depends on the treatment value (which
corresponds to the arrow from \(Z\) to \(h(M)\) in Figure
\ref{fig:fig1}(c)). For example in the ``current-slope'' case the slope
must be dependent on the treatment. This can be achieved by adding an
interaction between treatment and time in the longitudinal model.
Moreover, the shared random-effects link is not relevant in a mediation
analysis perspective as the random effects are assumed independent of
the covariates.

\subsubsection{Likelihoods and estimations}

Both Model (\ref{eq:model1}) and Model (\ref{eq:model2}) can be
estimated through maximum likelihood. The baseline hazard functions can
be estimated parametrically using M-splines which provides flexibility
in their estimations. A penalized likelihood can be maximized instead of
the full likelihood to ensure smooth estimate of these functions
\citep{joly1998penalized,rondeau2007joint}. Let \(\phi_1\) and
\(\phi_2\) be the vectors of all the parameters in Model
(\ref{eq:model1}) and Model (\ref{eq:model2}) respectively. The
penalized likelihood for Model (\ref{eq:model1}) is given by:
\begin{equation} \label{eq:likelihood1}
 L_{\text{pen}}(\phi_1) = L(\phi_1) - \kappa_T \int_0^{\infty} \left(\lambda^{''}_{0,T}(t)\right)^2 \mathrm{d}t -  \kappa_S \int_0^{\infty} \left(\lambda^{''}_{0,S}(t)\right)^2\mathrm{d}t
\end{equation} and for the Model (\ref{eq:model2}):

\begin{equation} \label{eq:likelihood2}
 L_{\text{pen}}(\phi_2) = L(\phi_2) - \kappa_T\int_0^{\infty} \left(\lambda^{''}_{0,T}(t)\right)^2 \mathrm{d}t
\end{equation} In these equations, the terms \(\kappa_S\) and
\(\kappa_T\) are penalization terms on the second derivatives of the
baseline hazard functions that ensure smooth estimations of these
functions.

In practice the maximization of the penalized likelihood is carried out
using the Levenberg-Marquardt algorithm \citep{marquardt1963algorithm}.

\subsection{Surrogacy measures}

In this section we provide definitions and formulas for the surrogacy
measures that can be used with each model. In the meta-analytic
approach, several measures have been proposed in the literature:
individual and trial-level associations and the surrogate threshold
effect (STE). From a mediation perspective the main measure of surrogacy
is the proportion of treatment effect (PTE), which quantifies the amount
of treatment effect on the final endpoint going through its effect on
the surrogate. In the following we present each of these measures. We
denote by \(\xi\) the vector of all random effects in Model
(\ref{eq:model1}) or Model (\ref{eq:model2}).

\subsubsection{Individual-level association}

Individual-level surrogacy is often represented by a measure of
association between the surrogate and the final endpoint, with the
Kendall's \(\tau\) between both endpoints being commonly used. We first
introduce the definition of the Kendall's \(\tau\) and then give a
derivation based on the two proposed joint models. Let \(X\) and \(Y\)
be two continuous random variables. For two random subjects
\((X_1,Y_1)\) and \((X_2,Y_2)\), Kendall's \(\tau\) is defined as the
probability that the two couples \((X_1,Y_1)\) and \((X_2,Y_2)\) are
concordant minus the probability that they are discordant. The couples
are concordant if \(X_1 > X_2\) and \(T_1>T_2\) or if \(X_1 < X_2\) and
\(T_1 < T_2\) which is equivalent to \((X_1-X_2)(Y_1-Y_2)>0\). They are
discordant otherwise. Therefore the Kendall's \(\tau\) for \((X,Y)\) can
be written as \citep{wang2000estimation}

\begin{align}\label{eq:ktau}
\begin{split}
\tau &= 4\mathbb P\left ( X_1> X_2 , Y_1 > Y_2  \right) -1.
\end{split}
\end{align} In the case where \(X\) and \(Y\) are independent,
\(\mathbb P\left ( (X_1> X_2 , Y_1 > Y_2 \right) =\mathbb P\left ( X_1> X_2 \right)\mathbb P\left ( Y_1> Y_2 \right)\)
and by symmetry
\(\mathbb P\left ( X_1> X_2 \right) = \mathbb P\left ( Y_1> Y_2 \right) = \frac{1}{2}\)
which gives \(\tau = 0\). On the other hand, for perfect correlation
\(X=Y\), \(\mathbb P\left (X_1> X_2 , Y_1 > Y_2 \right)\) =
\(\mathbb P\left ( X_1> X_2 \right) = \frac{1}{2}\) and then
\(\tau = 1\) while
\(X=-Y \Rightarrow \mathbb P\left (X_1> X_2 , Y_1 > Y_2 \right )= 0 \Rightarrow \tau =-1.\)
Hence, \(\tau\) takes values in \([-1,1]\) with values close to \(0\)
meaning small association between \(X\) and \(Y\).

\paragraph{(a) Kendall's $\tau$ for a time-to-event surrogate}

\mbox{} \newline   

For Model (\ref{eq:model1}) the Kendall's \(\tau\)
is equal to (for the variables \(S\) and \(T\)),
\begin{align}\label{eq:tau1}
\begin{split}
\tau = 4\int_{\mathbb R^4} \Bigg [ & \int_0^\infty \int_0^\infty \left( \int_0^\infty \mathbb P(T_1 > t \mid S_1 = s+\alpha,\xi_1) f_{T_2}(t \mid S_2=s,\xi_2) \mathrm{d}t \right) \\ 
& \times f_{S_1}(s+\alpha \mid \xi_1) f_{S_2}(s \mid \xi_2) \mathrm{d}s \mathrm{d}\alpha \Bigg ] \mathbb P(S_1 > S_2 \mid \xi_1,\xi_2) \mathrm{d}f(\xi_1,\xi_2) \xi_1 \mathrm{d}\xi_2-1.
\end{split}
\end{align}
In the case where there is no effect of \(S\) on \(T\)
(i.e.~independence between \(S\) and \(T\) given the random effects or
\(\gamma(S) =0\)), this expression reduces to
\begin{align}\label{eq:tau2}
\begin{split}
\tau = &\int_{\mathbb R^4}  \left [  \frac{e^{\omega_2 + u_2}+ e^{\zeta \omega_2 + \alpha u_2 }}{\left(e^{\omega_1+u_1}+ e^{\omega_2+u_2} \right)\left( e^{\zeta \omega_1 + \alpha u_1 } + e^{\zeta \omega_2 + \alpha u_2 }\right)} \right] \\ & \qquad \frac{1}{4\pi^2 \theta \gamma} \exp\left( -\frac{1}{2}\left( \frac{\omega_1+\omega_2}{\theta^2} + \frac{u_1+u_2}{\gamma^2}  \right) \right) \mathrm{d}\omega_1 \mathrm{d}\omega_1 \mathrm{d}u_1 \mathrm{d}u_2.
\end{split}
\end{align} which can be easily approximated using Monte Carlo sampling
\citep{sofeu2019one}.

\subsubsection{Trial-level association}

For both Model (\ref{eq:model1}) and Model (\ref{eq:model2}), the
trial-level association is defined as the coefficient of determination
of the random effects \((\nu_S, \nu_T)\) or \((\nu_M, \nu_T)\). For both
models these random effects are assumed to be Gaussian with mean \(0\)
and covariance matrix \(\Sigma_{\nu}\). If we write,

\[
\Sigma_\nu = \begin{pmatrix} 
\sigma_{\nu_1}^2 & \sigma_{\nu_{12}} \\
\sigma_{\nu_{12}} & \sigma_{\nu_2}^2
\end{pmatrix}.
\] Then, the trial-level measure of surrogacy, \(R^2_{\mathrm{trial}}\)
is defined as: \[
R^2_{\mathrm{trial}} = \frac{\sigma_{\nu_{12}}^2 }{\sigma_{\nu_1}^2\sigma_{\nu_2}^2}.
\]

As for the Kendall's \(\tau\), standard error for the estimator of
\(\hat R^2_{\mathrm{trial}}\) can be obtained using parametric
bootstrap. However, the simple form of \(R^2_{\mathrm{trial}}\) allows
the derivation of an estimator of its standard error directly from the
estimated covariance matrix of the vector of parameter using the
delta-method and then using the normal approximation to derive a
confidence interval. However the later could yield confidence intervals
that exceeds the range \([0,1]\) of \(R^2_{\mathrm{trial}}\).

\subsubsection{Surrogate threshold effect}

The surrogate threshold effect is the smallest treatment effect on the
surrogate endpoint required to yield a significant effect on the final
endpoint \citep{burzykowski2006surrogate}.

The prediction interval for the treatment effect on the final endpoint
in a new trial \(i_0\), \(\beta_{Z,T}+\nu_{T,i_0}\), given that the
observed effect on the surrogate is \(\beta_{Z,S}+\nu_{S,i_0}\) is:

\[
\mathbb E \left ( \beta_{Z,T}+\nu_{T,i_0} \mid \nu_{S,i_0} \right ) \pm z_{1-\gamma/2} \sqrt{\mathbb V\left( \beta_{Z,T}+\nu_{T,i_0} \mid \nu_{S,i_0}  \right)} \ ,
\] where \(z_{1-\gamma/2}\) is the \((1-\gamma/2)\) percentile of the
standard normal distribution. In order to predict a significant effect
of the treatment on the final endpoint, the lower bound of this interval
should be greater than \(0\) (or the upper bound lower than \(0\)
depending on the type of effect one is interested in). This interval
only depends on \(\nu_{S,i_0}\). Let \(l(\nu_{S,i_0})\) be the lower
bound of this interval, then the STE is the value of \(\nu_{S,i_0}\)
such that \[
l(\nu_{S,i_0}) = 0. 
\]
\subsubsection{Proportion of treatment effect}

The PTE is a measure related to mediation analysis. Mediation analysis
investigates how the total effect of a treatment on a outcome can be
decomposed as a direct effect and an indirect effect through a mediator
(intermediate variable). Most development of mediation analysis have
been made in the realm of causal inference, mostly based on
counterfactual outcomes. In our case, the mediator is the surrogate
endpoint of interest and the endpoint is the final endpoint. Let, \[
\mathbb S^{zz'}(t) = \mathbb P(T(z,S(z') > t),
\] be the survival function of the final endpoint \(T\) if the treatment
for \(T\) is set to \(z\) but to \(z'\) for the surrogate \(S\) (here we
do not make a distinction between a time-to-event surrogate or a
longitudinal surrogate). Details on the interpretations of the above
formula have been published \cite{lecoent2022time}.

For a binary treatment we can derive its natural indirect effect (NIE)
on \(T\) through \(S\) as: \[
\nie(t)  = \mathbb S^{11}(t)  - \mathbb S^{10}(t).
\] The righthand side of this formula is the difference between the
survival function of \(T\) in the case where the treatment is set to
\(1\) for both \(S\) and \(T\) and the survival function of \(T\) in the
case where the treatment is set to \(1\) for \(T\) but to \(0\) for
\(S\). The only difference between \(\mathbb S^{11}(t)\) and
\(\mathbb S^{10}(t)\) is the treatment value for \(S\), hence their
difference quantifies the amount of treatment effect (on the survival
scale) on \(T\) due to its effect on \(S\). In the same manner we can
define the natural direct effect as: \[
\nde(t)  = \mathbb S^{10}(t)  - \mathbb S^{00}(t).
\] The total treatment effect (TE) is then the sum of the NIE and NDE,
\(\tte(t) = \nie(t) + \nde(t)\). Finally, the measure of surrogacy is
defined as: \begin{equation} \label{eq:pte}
\begin{split}
\pte(t) &= \frac{\nie(t)}{\tte(t)} \\
&= \frac{\mathbb S^{11}(t)  - \mathbb S^{10}(t)}{\mathbb S^{11}(t)  - \mathbb S^{00}(t)}.  
\end{split}
\end{equation} which is the proportion of treatment effect going through
the surrogate. To be able to compute the function \(\pte(t)\) we need to
compute the functions \(\mathbb S^{zz'}(t)\), which can be done from
Model (\ref{eq:model1}) and Model (\ref{eq:model2}) using the formula
\citep{lecoent2022time,zheng2021quantifying},
\begin{equation}\label{eq:szz'}
\begin{split}
&\mathbb S^{zz'}(t) = \int_{\mathcal{X}}\int_0^\infty  \int_{\xi} \mathcal S_T(t \mid s, z, \xi,X=x) f_S(s \mid z', \xi,x) f_\xi(\xi) f_X(x) \text{d}s  \text{d}\xi  \text{d}x \quad \text{for Model (\ref{eq:model1})} \\
& \mathbb S^{zz'}(t) =  \int_{\mathcal{X}} \int_\xi \mathcal S_T\left(t \mid h(M^{z'}),\xi,z,x\right) f_\xi(\xi) f_X(x)  \text{d}\xi \text{d}x \quad \text{for Model (\ref{eq:model2}),}
\end{split}
\end{equation} where \(f_\xi(\cdot)\) and \(f_X(\cdot)\) are the density
functions of the random-effects and covariates in the models
respectively. In the first equation, \(S_T(t \mid s, z', \xi,x)\) is the
survival function of \(T\) evaluated at \(t\), given the treatment is
set to \(z\), the surrogate \(S=s\), the random effects \(\xi\) and
covariates \(x\). The function \(f_S(s \mid z, \xi,x)\) is the density
of \(S\) given the treatment is set to \(Z=z'\), the random effect and
covariates. In the second equation,
\(S_T\left(t \mid h(M^{z'}),\xi,Z=z,X=x\right)\) is the survival
function of \(T\) given the treatment \(Z=z\), the random effects \(x\),
the covariates \(x\) and the link function \(h(M^{z'})\) given the
treatment for \(M\) is set to \(z'\). Note that in the second case, in
contrary to the first case, we do not need to integrate over the
distribution of \(M^{z'}\) since it only depends on \(z'\), the random
effects \(\xi\), and the covariates \(x\). From Equation \ref{eq:szz'},
\(\pte(t)\) can be obtained by setting the appropriate values for \(z\)
and \(z'\) following Equation \ref{eq:pte}. All the functions involved
in the righthand side of Equation \ref{eq:szz'} can be derived from the
estimated parameters of the model, except \(f_X(\cdot)\) for which
averaging over the observed distribution of the covariates in the
dataset can be used to approximate the integral over \(\mathcal X\). As
for the Kendall's \(\tau\) and \(R^2_{\mathrm{trial}}\), the standard
error and confidence interval of \(\pte(t)\) can be obtained through
parametric bootstrap.

\subsection{Implementation in frailtypack}

In this Section we present two main functions of the
\verb|frailtypack| package for investigating surrogacy and their main
arguments. A detailed explanation of these functions can be found in the
reference manual of the package available in R or on the CRAN \citep{frailtypackcite}. In this article we use \verb|frailtypack| version 3.6.5 and R version 4.4.0.

\subsubsection{Function jointSurroPenal}

The function \verb|jointSurroPenal| can be used to investigate
surrogacy when both the surrogate \(S\) and the final endpoint \(T\) are
time-to-event (model (\ref{eq:model1})). The call to this function is as
follows (the values given for each parameters are the default values):

\begin{small}
\begin{verbatim}
R> model<- jointSurroPenal(data, maxit = 50,
+          indicator.zeta = 1, indicator.alpha = 1, 
+          frail.base = 1, n.knots = 6, LIMparam = 0.001, 
+          LIMlogl = 0.001, LIMderiv = 0.001, nb.mc = 300,
+          nb.gh = 32, nb.gh2 = 20, adaptatif = 0, int.method = 2, 
+          nb.iterPGH = 5, nb.MC.kendall = 10000, nboot.kendall = 1000, 
+          true.init.val = 0, theta.init = 1, sigma.ss.init = 0.5, 
+          sigma.tt.init = 0.5, sigma.st.init = 0.48, 
+          gamma.init = 0.5, alpha.init = 1, 
+          zeta.init = 1, betas.init = 0.5, betat.init = 0.5, 
+          scale = 1, random.generator = 1, kappa.use = 4, 
+          random = 0, random.nb.sim = 0, 
+          seed = 0, init.kappa = NULL, 
+          ckappa = c(0,0), nb.decimal = 4, 
+          print.times = TRUE, print.iter = FALSE, 
+          mediation=FALSE, g.nknots=1, 
+          pte.times=NULL, pte.ntimes=NULL,
+          pte.nmc=500, pte.boot=FALSE, pte.nboot=2000, 
+          pte.boot.nmc=500, pte.integ.type=2)
\end{verbatim}
\end{small}
In order to use this function, the user has to provide a dataset
(argument \verb|data|) of the following structure: 
\begin{small}
\begin{verbatim}
R> head(data)
\end{verbatim}
\end{small}
\begin{small}
\begin{verbatim}
patientID    timeT    timeS statusT statusS trt trialID
        1 9.057946 2.217739       1       1   0       1        
        2 2.986813 1.389263       1       1   0       1
        3 8.874237 8.874237       1       0   1       1 
        4 3.245388 1.809671       1       1   1       1
        5 4.448964 2.603604       1       1   0       1
\end{verbatim}
\end{small}
The dataset must contain one line per subject and seven columns: one for
the subject's identification number (column \verb|patientID|), for the
trial number (\verb|trialID|), treatment indicator (\verb|trt|)as
well as one for the follow-up time for the surrogate (\verb|timeS|)
and censoring indicator (\verb|statusS|) and for the follow-up time
for the final endpoint (\verb|timeT|) and censoring indicator
(\verb|statusT|).

The arguments \verb|indicator.zeta| and \verb|indicator.alpha| are
used to indicate if one wants to estimate the parameters \(\zeta\) and
\(\alpha\) in Model (\ref{eq:model1}). If they are set to \(0\) then
these parameters will not be estimated and assumed to be equal to \(1\),
which means that the random effects will have the same effect on \(S\)
and \(T\). The parameter \verb|frail.base| indicates if the user wants
to include the trial-level random effects \(u_i\) in the model. If set
to \(0\) then these random effects are not included, and neither will be
the parameter \(\alpha\) associated with \(u_i\). The number of knots
used in the M-splines basis for estimating the functions
\(\lambda_{0,S}(\cdot)\) and \(\lambda_{0,T}(\cdot)\) can be set through
the argument \verb|n.knots|, with allowed values between \(4\) and
\(20\) knots.

For the parametric bootstrap used to derive the standard error and
confidence interval for the estimated Kendall's \(\tau\), the argument
\verb|nboot.kendall| can be used to fix the number of bootstrap
samples to be generated.

The option to include the function \(\gamma(S)\) in Model
(\ref{eq:model1}) is given by setting the argument \verb|mediation| to
\verb|TRUE|. In that case the function \(\gamma(S)\) is estimated
using a basis of B-splines whose number of knots is given by the
argument \verb|g.nknots|, which can take any value between \(1\) and
\(5\) and the function \(\pte(t)\) will also be estimated. The
timepoints at which this function has to be evaluated can be specified
through the argument \verb|pte.times|. Note that these times should be
between \(0\) and the maximum observed final endpoint times. If one does
not want to specify any timepoints, the argument \verb|pte.ntimes| can
be used instead to specify the number of timepoints at which \(\pte(t)\)
should be evaluated. These points will then be selected evenly on the
range of the observed event times. The argument \verb|pte.boot| is
used if we want to compute quantile-based confidence bands of
\(\hat \pte(t)\) using parametric bootstrap. If set to \verb|TRUE|,
then the number of bootstrap samples to be used can be set with
\verb|pte.nboot|.

Other parameters are mainly pertain to computational details. A complete
description of each parameter can be found in the documentation of the
package \verb|frailtypack| available on the \verb|CRAN|.

The function \verb|jointSurroPenal| returns an \verb|R| object of
class \verb|jointSurroPenal| if the argument \verb|mediation| is set
to \verb|FALSE| and of class \verb|jointSurroMed| otherwise. In both
cases, common \verb|R| functions such as \verb|summary|,
\verb|print| and \verb|plot| can be used as will be illustrated in
Section 4.

\subsubsection{Function longiPenal}

The function \verb|longiPenal| can be used to investigate surrogacy
when the surrogate outcome is a longitudinal biomarker and the final
endpoint is a time-to-event (Model (\ref{eq:model2})). The call to this
function is as follows (the values given to each parameters are the
default values):

\begin{small}
\begin{verbatim}
R> model<- longiPenal(formula, formula.LongitudinalData, 
+           data,  data.Longi, formula.Binary = FALSE, 
+           random, random.Binary = FALSE, 
+           fixed.Binary = FALSE, GLMlog = FALSE, 
+           MTP = FALSE, id, intercept = TRUE, 
+           link = "Random-effects", timevar = FALSE, 
+           left.censoring = FALSE, n.knots, 
+           kappa, maxit = 350,  hazard = "Splines", 
+           mediation = FALSE, med.center = NULL, med.trt = NULL, 
+           init.B, init.Random, init.Eta, 
+           method.GH = "Standard", seed.MC = 1, n.nodes, 
+           LIMparam = 1e-3, LIMlogl = 1e-3, LIMderiv = 1e-3, 
+           print.times = TRUE,  med.nmc = 500, pte.times = NULL, 
+           pte.ntimes = NULL, pte.nmc = 500, 
+           pte.boot=FALSE,pte.nboot=2000)
\end{verbatim}
\end{small}
This function requires the specification of two datasets. The first one,
specified through the argument \verb|data|, contains the data
regarding the final endpoint \(T\) such as the follow-up time for each
subject, the censoring indicator, and potential baseline covariates.
Note that this dataset requires one line per subject and therefore does
not allow for time-dependent covariates to be included. Associated with
this dataset is the \verb|formula| wich is a formula object, with the
response on the left of a \(\sim\) operator, and the covariates on the
right. The response must be a survival object as returned by the
\verb|Surv| function of the R \verb|survival| package \citep{survival-package}. The variables
used in \verb|formula| should be the ones contained in \verb|data|.
For the longitudinal part, the repeated measurements data are specified
in a separate dataset through the argument \verb|data.Longi|. The
specification of the longitudinal submodel is made through \newline
\noindent \verb|formula.LongitudinalData| which is a \verb|R|
formula with the observed biomarker on the left and the different
covariates on the right. Both the names for the biomarker and the
covariates specified in this formula should correspond to columns in the
dataset \verb|data.Longi|.

Both \verb|data| and \verb|data.Longi| should have a column labelled
``\verb|id|'' that corresponds to the identificator of each subject in
order to link the two datasets, i.e., \verb|id=1| in \verb|data|
should corresponds to the same individual with \verb|id=1| in
\verb|data.Longi|. Note that for simpliciy the variable \verb|id|
should takes values between \(1\) and \(n_i\) where \(n_i\) is the total
number of subjects.

The argument \verb|intercept| can be used to indicate if one wants to
include the fixed intercept \(\beta_0\) in the longitudinal submodel,
with default value being \verb|TRUE|.

The argument \verb|timevar| is used to specify the timepoints of the
repeated measurements in \verb|data.Longi| for each
individual. This variable can be included in
\verb|formula.LongitudinalData| but is not required, a transformation
of \verb|timevar| can be used instead to take into account non linear
time evolution. However, in both case a variable \verb|timevar| must
be included in the dataset which contains the measurement times.

The individual-level random effects to be included in the longitudinal
submodel are specified through the argument \verb|random| which must
be a vector of characters corresponding to covariates in \verb|data.Longi|. 
The character \verb|"1"| is used to indicate a random intercept. For example if one wants to include
a random intercept and a random slope then one should specify
\verb|random=c("1","timevar")| (given that \verb|timevar| has been
included in \verb|formula.LongitudinalData|).

The baseline hazard function estimation type is chosen via the argument
\verb|hazard|. Possible values are \verb|"Splines"| and
\verb|"Splines-per"| for flexible hazard functions estimated with
M-Splines with respectively equidistant or percentile knots, or
\verb|"Weibull"| to specify a parametric Weibull hazard function. For
\verb|"Splines"| and \verb|"Splines-per"| the number of knots used
is set through the argument \verb|n.knots|.

The mediation analysis is enabled by setting the argument
\verb|mediation| to \verb|TRUE|. In that case one should also
specify the name of the variable in \verb|data| that corresponds to
the treatment through the argument \verb|treatment|. Moreover, one
should also specify the centers (or trials) to which each subject
belongs through the argument \verb|centers|. This argument takes a
character string which must correspond to the name of the variable in
\verb|data| indicating the center/trial of each subject. Note that as
the variable \verb|id|, the variable indicating the center for each
subject should take values between \(1\) and \(K\) where \(K\) is the
total number of centers/trials. If \verb|mediation| is set to
\verb|TRUE| then the function \(\pte(t)\) will be estimated. As for
the function \verb|jointSurroPenal|, one can specify the timepoints at
which \(\hat \pte(t)\) should be evaluated or the number of timepoints
at which it should be evaluated. The argument \verb|pte.boot| takes
values \verb|TRUE/FALSE| to indicate if the parametric bootstrap
estimation of the standard-error of \(\hat \pte(t)\) and its confidence
bands should be computed. If set to \verb|TRUE| then the number of
bootstrap samples is specified by \verb|pte.nboot| with a default
value of \(2000\). A complete description of each parameter can be found
in the documentation of the package \verb|frailtypack| available on
the \verb|CRAN|.

The function \verb|longiPenal| returns a \verb|R| object of class
\verb|longiPenal| on which the usual \verb|R| functions
\verb|summary|, \verb|print| and \verb|plot| can be applied as
will be illustrated in Section 4.

\section{Illustrations}

We illustrate the proposed methods in three applications on cancer data
from meta-analyses or multicentric randomized clinical trial. The first
application is based on a dataset on ovarian cancer, the second on
gastric cancer and the third on colorectal cancer. In the following we assume that the \verb|frailtypack| package is loaded using the \verb|R| commands \verb|require(frailtypack)| or \verb|library(frailtypack)|.

\subsection{Progression-free survival as a surrogate of overall survival in advanced ovarian cancer: meta-analytic approach}

In this first application we are interested in evaluating
progression-free survival as a surrogate endpoint for overall survival
in advanced ovarian cancer using the classical meta-analytic approach
and its two-level validation. The dataset is the \verb|dataOvarian|
which can be loaded from \verb|frailtypack| using the following
command:

\begin{small}
\begin{verbatim}
R>  data("dataOvarian")
\end{verbatim}
\end{small}
This dataset combines the data of four double-blind randomized clinical
trials. These trials investigated the efficacy of cyclophosphamide plus
cisplatin versus cyclophosphamide plus adriamycin plus cisplatin in
advanced colorectal cancer \citep{ovarian1991cyclophosphamide}. The
structure of the dataset is as follows:

\begin{small}
\begin{verbatim}
R> head(dataOvarian)
\end{verbatim}
\end{small}
\begin{small}
\begin{verbatim}
  patientID trialID trt      timeS statusS     timeT statusT
1         1       2   0 0.10515873       1 0.1857143       1
2         2       2   0 0.89523809       1 1.4087302       1
3         3       2   0 0.07896825       1 0.1261905       1
4         4       2   1 1.73928571       0 1.7392857       0
5         5       2   0 0.09126984       1 0.1273810       1
6         6       2   1 0.16984127       1 0.2253968       1
\end{verbatim}
\end{small}
\normalsize The columns \verb|patientID| and \verb|trialID|
indicate the identifiers of the subject and the trial respectively. The
variable \verb|trt| takes value \(0\) for cyclophosphamide plus
cisplatin and \(1\) for cyclophosphamide plus adriamycin and cisplatin.
The variables \verb|timeS| and \verb|statusS| are the follow-up and
censoring indicator of the time-to-progression (the surrogate) while
\verb|timeT| and \verb|statusT| are the follow-up and censoring
indicator of the time-to-death (from any cause).

\subsubsection{Model fitting and surrogacy evaluation}

Once the dataset has been loaded, we can call the
\verb|jointSurroPenal| function as follows:

\begin{small}
\begin{verbatim}
R> mod.ovar <- jointSurroPenal(data = dataOvarian, n.knots = 8,
+              indicator.alpha = 0, nb.mc = 500, scale = 1/365)
\end{verbatim}
\end{small}
Here we do not specify the option \verb|mediation=TRUE| (the default
value being \verb|FALSE|), therefore the approach used corresponds to
a ``classical'' meta-analytic approach rather than a mediation approach
and is based on Model (\ref{eq:model1}) without inclusion of the
function \(\gamma(S)\).

We specify that the number of knots (inner + boundary) in the M-splines
basis for estimating the two baseline hazard functions \(\lambda_{0,S}\)
and \(\lambda_{0,T}\) is \(8\) (\verb|n.knots = 8|). The argument
\verb|indicator.alpha = 0| indicates that we don't estimate the
parameter \(\alpha\) which will be fixed to \(1\) throughout the
estimation procedure (see Model (\ref{eq:model1})). The argument
\verb|nb.mc = 500| specifies that the number of Monte Carlo points
used in the approximation of the integral on the trial-level random
effects is set to \(500\). Finally \verb|scale = 1/365| is used to
rescale the follow-up times in days (originally expressed in years).
The object \verb|mod.ovar| belongs to the class
\verb|jointSurroPenal| on which we can apply the R \verb|summary|
function.
The results of the estimation can be simply outputed using the command
\verb|print(mod.ovar)|. The base \verb|R| function can be used \verb|summary| to displays estimators of the estimated fixed treatment effects and surrogacy measures.

\begin{small}
\begin{verbatim}
R> print(mod.ovar)
\end{verbatim}
\end{small}
\begin{small}
\begin{verbatim}
Estimates for variances parameters of the random effects 
         Estimate Std Error      z       P    
theta       6.852    0.3781 18.121  < e-10 ***
zeta        1.788    0.0709 25.224  < e-10 ***
gamma       0.062    0.0855  0.724  0.4691    
sigma2_S    0.514    0.3375  1.522  0.1279    
sigma2_T    1.572    0.9249  1.700 0.08918   .
sigma_ST    0.899    0.5502  1.634  0.1023    
  
Estimates for the fixed treatment effects 
       Estimate Std Error      z        P   
beta_S   -0.696    0.2239 -3.108 0.001883 **
beta_T   -1.017    0.3824 -2.660 0.007813 **
--- 
Signif. codes:  0 '***' 0.001 '**' 0.01 '*' 0.05 '.' 0.1 ' ' 1  
  
hazard ratios (HR) and confidence intervals for the fixed treatment effects 
       exp(coef) Inf.95.CI Sup.95.CI
beta_S     0.499     0.322     0.773
beta_T     0.362     0.171     0.765
  
Surrogacy evaluation criterion 
             Level Estimate Std Error Inf.95.CI Sup.95.CI Strength
Ktau    Individual    0.681        --     0.663     0.693         
R2trial      Trial    1.000     0.002     0.997     1.003     High
R2.boot      Trial    0.975        --     0.847     1.000     High
--- 
Correlation strength: <= 0.49 'Low'; ]0.49 - 0.72[ 'Medium'; >= 0.72 'High'  
--- 
The treatment effects on the surrogate endpoint (beta_S) 
that can predict a nonzero  treatment effect on the true endpoint (beta_T) 
belongs to the interval: ]-Inf ; -0.296[ : HR= ]0 ; 0.744[
Surrogate threshold effect (STE) : -0.296 (HR = 0.744 ) 
  
Convergence parameters 
Penalized marginal log-likelihood =  -10892.959 
Number of iterations =  22 
Smoothing parameters =  60093516.4969293 0.000207706120839778 
Number of spline nodes =  8 
LCV = the approximate likelihood cross-validation criterion 
      in the semi parametrical case     =  9.162 
Convergence criteria: 
Parameters =  2.198e-06 Likelihood =  8.474e-06 Gradient =  8.65e-10 
Estimation based on a combination of 
both Gaussian-Hermite quadrature and Monte Carlo integration
\end{verbatim}
\end{small}
The first part of the results shows the estimated variance of the
individual and trial-level random effects and the power parameters
associated with \(\omega_{ij}\) and \(u_i\) in the hazard function of
\(T\) (respectively \(\alpha\) and \(\zeta\)). Note that here we only
have an estimation for \(\zeta\) since we specified
\verb|indicator.alpha = 0|. These results show a significant variance of
the individual-level random effect \(\omega_{ij}\) (variable
\verb|theta|). Since the parameter \(\zeta\) is also significantly
different from \(0\), both \(S\) and \(T\) are positively correlated at
the individual level. On the other hand, the trial-level random effects
variance (variable \verb|gamma|) are not significantly different from \(0\). The variables
\verb|sigma2_S|, \verb|sigma2_T| and \verb|sigma_ST| are respectively
the estimations of the variance of \(\nu_{S,i}\), the variance of
\(\nu_{T,i}\) and their covariance.

The second part of the results show the estimated fixed treatment
effects, their estimated standard errors and the associated Wald
statistics and p-values. There are significant treatment effects on both
endpoints, therefore the addition of adriamycin to cyclophosphamide plus
cisplatin reduces the risk of the occurrences of progression and death.

The third part of the results shows the hazard-ratios associated with
the estimated fixed treatment together with their \(95\%\) confidence
intervals, with a hazard ratio of \(0.499\) (\([0.322, 0.773]\)) for the
surrogate and \(0.362\) (\([0.171, 0.765]\)) for the final endpoint.

The fourth part of the results shows the estimated surrogacy measures at
the individual and trial level. The variable \verb|Ktau| is the
estimated Kendall's \(\tau\) between the surrogate and the final
endpoint with its associated \(95\%\) confidence interval. The estimated
value of \(0.681\) suggests a mild association between the two
endpoints. The variable \verb|R2trial| is the estimated
\(R^2_{\text{trial}}\) with its estimated standard error (obtained using
the delta-method) and its \(95\%\) confidence interval while
\verb|R2.boot| obtained using parametric bootstrap. Their values close
to \(1\) suggest a high association between the treatment effects on the
surrogate and the final endpoint at the trial-level. Finally these
results suggest that the time-to-progression can be considered as a good
surrogate of the overall survival in this setting of advanced ovarian
cancer.

Additionally, the results give the surrogate threshold effect estimated
at \(-0.296\) (associated hazard-ratio \(0.744\)). This means that in a
new trial, the minimal treatment effect on the surrogate to be observed
in order to predict a significant treatment effect on the final endpoint
is \(-0.296\).

Moreover, the estimated baseline hazard functions \(\lambda_{0,S}\) and
\(\lambda_{0,T}\) can be plotted from the \verb|mod.ovar|
object using the \verb|R| function \verb|plot| from the following command:

\begin{small}
\begin{verbatim}
R>  plot(mod.ovar, pos.legend = "topleft")
\end{verbatim}
\end{small}

\begin{center}\includegraphics[width=0.7\linewidth]{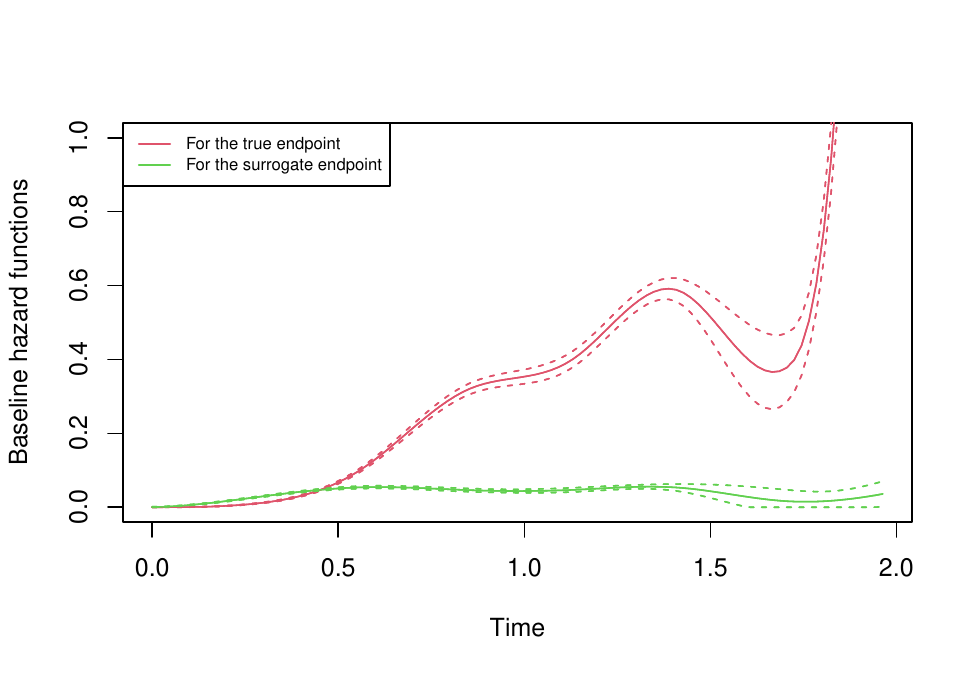} \end{center}

It should be noted that the results in this application have to be
interpreted with caution. The broad confidence interval and a point estimate close to  \(1.0\) for \(R^2_{\text{trial}}\) may indicates convergence issues, even if the three convergence criteria (based on the likelihood, its first order derivative and its second order derivative) show that the model converged.

\subsection{Time-to-relapse as a surrogate of overall survival using proportion of treatment effect in gastric cancer: a mediation approach}

The second application is on a meta-analysis on resectable gastric
cancer patients investigating the addition of adjuvant chemotherapy
after surgery versus surgery alone \citep{paoletti2010benefit}. In this
illustration, the final endpoint is the time between randomization and
death from any cause while the surrogate is the time-to-relapse, defined
as the time between randomization and disease recurrence or occurrence
of a second cancer, whichever occurred first. Therefore both endpoints
might be right censored due to loss to follow-up and moreover the
surrogate endpoint might be censored by the final endpoint. We are
interested in estimating the proportion of treatment effect (adjuvant
chemotherapy or not) on overall survival that goes through its effect on
time-to-relapse.

\subsubsection{Dataset}

As for the first illustration, the dataset \verb|gastadj| can be
loaded directly from \verb|frailtypack| using the command

\begin{small}
\begin{verbatim}
R>  data("gastadj")
\end{verbatim}
\end{small}

This dataset contains the data of \(3288\) patients from \(14\)
randomized clinical trials. Out of these \(3288\) patients, \(1654\)
were assigned to the control group of no adjuvant chemotherapy, and the
remaining \(1634\) patients were assigned to receive adjuvant
chemotherapy. The dataset has the following structure, using the command
\verb|head(gastadj)|,

\begin{small}
\begin{verbatim}
R>  head(gastadj)
\end{verbatim}
\end{small}
\begin{small}
\begin{verbatim}
trialID patientID trt timeT statusT timeS statusS
1       1         1   1  4636       0  4636       0
2       1         2   1  4536       0  4536       0
3       1         3   0  3151       1  3151       1
4       1         4   1   485       1   432       1
5       1         5   0   435       1   300       1
6       1         6   0   187       1   137       1
\end{verbatim}
\end{small}
The columns \verb|trialID, patientID, trt| are the trial, patient and
treatment indicator respectively. The variables \verb|timeT, timeS|
are the follow-up times for the final endpoint and surrogate endpoints
respectively and \verb|statusT|, \verb|statusS| their associated
censoring indicator.
In this dataset, the variable \verb|timeS| corresponds to a
time-to-progression defined as the earliest between cancer recurrence,
occurrence of a second cancer or death. Therefore this endpoint includes
death as a composite endpoint which raises questions from a mediation
analysis viewpoint since the final endpoint always triggers the
surrogate. To circumvent this, we instead analyzed the time-to-relapse
(cancer recurrence or second cancer) by censoring them at the time of
death. In the dataset this change can be made using the following
command,

\begin{small}
\begin{verbatim}
R>  gastadj[gastadj$timeS == gastadj$timeT &
+   gastadj$statusS == 1, c("statusS")] <- 0
\end{verbatim}
\end{small}
For practical purposes, and to reduce the computation time, we restrain
this illustration on a subset of the original dataset, by selecting
\(20\%\) of the patients at random.

Moreover, to circumvent some computing issues, we divide the time
variable (originally represented in a daily scale) by \(365\) in the
yearly scale. Therefore, the full call for data preparation is

\begin{small}
\begin{verbatim}
R>  data(gastadj)
R>  gastadj$timeS <- gastadj$timeS/365
R>  gastadj$timeT <- gastadj$timeT/365
R>  #"statusS" corresponds now
R>  #to a time-to-relapse event
R>  gastadj[gastadj$timeS == gastadj$timeT &
R>  gastadj$statusS == 1, c("statusS")] <- 0
R>  # select 20% of the original dataset
R>  set.seed(1)
R>  n <- nrow(gastadj)
R>  subset <- gastadj[sort(sample(1:nrow(gastadj),
+             round(n*0.2), replace = F)),]
\end{verbatim}
\end{small}

\subsubsection{Model fitting and surrogacy evaluation}

The call to the function \verb|jointSurroPenal| is the following:

\begin{small}
\begin{verbatim}
R>  mod.gast<-jointSurroPenal(subset,n.knots = 4,
+             indicator.zeta = 0, indicator.alpha = 0,
+             mediation = TRUE, g.nknots = 1,
+             pte.times = seq(1.5, 2,length.out = 30),
+             pte.nmc = 10000, pte.boot = TRUE, pte.nboot = 1000,
+             pte.boot.nmc = 1000)
\end{verbatim}
\end{small}
In this call we set both \verb|indicator.zeta| and
\verb|indicator.alpha| to \(0\), therefore the parameters \(\zeta\)
and \(\alpha\) in Model (\ref{eq:model1}) are not estimated and assumed
to be equal to \(1\).

Here we specify that \verb|mediation = TRUE|, therefore the function
\(\gamma(S)\) in Model (\ref{eq:model1}) will be estimated using
B-Splines. The number of inner knots used in the spline basis is fixed
to \(1\) via the command \verb|g.nknots=1|. Since we are interested in
the mediation analysis setting, we specify that we want the function
\(\pte(t)\) to be evaluated at \(30\) timepoints defined by the argument
\verb|pte.times|. The number of Monte-Carlo points used in
the approximation of the integral over the random effects in Equation
\ref{eq:szz'} is set to 10000. The use of parametric bootstrap to derive
estimated standard-errors and confidence bands for \(\pte(t)\) is given
by \verb|pte.boot=TRUE| where we also specify that we want this
bootstrap to be based on \(1000\) sampling via \verb|rt.nboot=1000|.
Finally, for illustration purposes and to reduce computation time, we
also set the number of Monte Carlo points used for each bootstrap sample
to \(1000\). However, in practice this number should be the same as for
the estimation of \(\pte(t)\). The object \verb|mod.gast| has R class
\verb|jointSurroMed|, and we can apply the \verb|summary| function
to display the results.

\begin{small}
\begin{verbatim}
R>  summary(mod.gast)
\end{verbatim}
\end{small}
\begin{small}
\begin{verbatim}
Results of a joint surrogate mediation model using a penalized likelihood
on the baseline hazard functions.
  
Estimates for variances parameters of the random effects 
         Estimate Std Error     z        P    
theta       5.061    0.5798 8.729   < e-10 ***
gamma       2.143    0.8059 2.659 0.007837  **
sigma2_S    0.551    0.3529 1.561   0.1184    
sigma2_T    0.601    0.3990 1.507   0.1317    
sigma_ST    0.575    0.3568 1.613   0.1068    
  
Estimates for the fixed treatment effects 
       Estimate Std Error      z      P  
beta_S   -0.404    0.3216 -1.257 0.2086  
beta_T   -0.280    0.3384 -0.827 0.4083  
--- 
Signif. codes:  0 '***' 0.001 '**' 0.01 '*' 0.05 '.' 0.1 ' ' 1  
  
Hazard ratios (HR) and confidence intervals for the fixed treatment effects 
       exp(coef) Inf.95.CI Sup.95.CI
beta_S     0.667     0.355     1.253
beta_T     0.756     0.389     1.467
  
Individual and trial level associations 
             Level Estimate Std Error Inf.95.CI Sup.95.CI Strength
Ktau    Individual    0.618        --     0.574     0.660         
R2trial      Trial    0.999     0.013     0.974     1.025     High
R2.boot      Trial    0.926        --     0.494     1.000     High
--- 
Correlation strength: <= 0.49 'Low'; ]0.49 - 0.72[ 'Medium'; >= 0.72 'High'  
--- 
  
Estimated function g at 3 times of occurence of the surrogate
    Time     g         CI.95
1  4.167 3.088 [2.567;3.609]
2  8.341 2.661 [1.483;3.838]
3 12.514 2.600 [0.185;5.015]
  
Estimated function PTE(t), natural direct, indirect and total effect 
at 3 time points 
   Time   PTE      CI.95.PTE   
1 1.655 0.281 [-0.292;0.457] 
2 1.828 0.273 [-0.003;0.573] 
3 2.000 0.042 [-0.308;0.444]

   Time     TE      CI.95.TE 
1 1.655  0.023 [0.015;0.031]
2 1.828  0.027 [0.019;0.038] 
3 2.000  0.026 [0.017;0.037] 

   Time    NDE     CI.95.NDE  
1 1.655  0.017 [0.012;0.029] 
2 1.828  0.020 [0.011;0.029] 
3 2.000  0.025 [0.014;0.034]

   Time    NIE      CI.95.NIE
1 1.655  0.007 [-0.005;0.012]
2 1.828  0.008      [0;0.017]
3 2.000  0.001 [-0.007;0.013]
  
Convergence parameters 
Penalized marginal log-likelihood =  -1362.853 
Number of iterations =  14 
LCV = the approximate likelihood cross-validation criterion 
      in the semi parametrical case     =  2.106 
Convergence criteria: 
  parameters =  0.0004349 likelihood =  0.0004285 gradient =  5.591e-07 
\end{verbatim}
\end{small}
As in the first illustration, we first have the estimation of the random
effects variances. The parameters \verb|theta| and \verb|gamma| (the
variances of \(\omega_{ij}\) and \(u_i\) in Model (\ref{eq:model1})
respectively) are significantly different from \(0\); there is an
association between \(S\) and \(T\) at the individual-level (through
\(\omega_{ij}\)) and trial-level (\(u_i\)). However, the variances of
\(\nu_S\) and \(\nu_T\) associated with the heterogeneity of the
treatment effects between trials are not significantly different from
\(0\).

The estimated fixed treatment effect on the surrogate and the final
endpoint are respectively \(-0.404\) and \(-0.280\). Given their large estimated standard-error, these effects are not significant.

Regarding the association measures of surrogacy we have an estimated
Kendall's \(\tau\) of \(0.618\) with a 95\% confidence interval
\([0.574,0.660]\) which suggests a poor individual-level surrogacy. At
the trial-level the estimated \(R^2_{\text{trial}}\) is \(0.999\) and its
95\% confidence interval is \([0.974,1.025]\). The
bootstrapped estimation and confidence interval are \(0.926\) and
\([0.494, 1.000]\), which suggest not enough evidence to support a good
trial-level surrogacy.

The next part of the results shows the value of the estimated function
\(\gamma(\cdot)\) at three timepoints. We can see that for the three
times the estimated values are significantly different from \(0\), which
suggest that there is a direct link between the occurrence of the
surrogate \(S\) and the risk of occurrence of \(T\). Moreover the
estimated function \(\gamma(S)\) being positive translates as an
increased risk of occurrence of \(T\) once the surrogate occurs.

The next part of the results displays the estimated values of the
function \(\pte(t)\) at the same three timepoints, as well as the
estimated total treatment effect, \(\tte(t)\), natural indirect and
direct effects, \(\nie(t)\) and \(\nde(t)\).

First, for \(\pte(t)\), we can see that for the three timepoints,
the estimated values of \(\pte(t)\) are small, with values ranging from $0.281$  
to $0.042$, with their confidence interval containing the value $0$. 
This suggest a poor proportion of treatment effect. Moreover, the wide confidence intervals
are due to a  total treatment effect, \(\tte(t)\), very close to $0$, therefore \(\pte(t)\)
defined as the ratio of \(\nie(t)\) over \(\tte(t)\) may be extremely
instable if \(\tte(t) \approx 0\), resulting in the wide confidence
interval. Both the direct effect, \(\nde(t)\), and indirect effect,
\(\nde(t)\), are also very close to $0$. 
The mediation measures of surrogacy, essentially \(\pte(t)\) and
associated \(\nie(t)\) and \(\nde(t)\), are inherently time-dependent.
Therefore a graphical representation might be of interest when
investigating surrogacy through theses measures. Especially the
\verb|plot| function can be applied to the object \verb|mod.gast| to
display the estimated results.
\begin{small}
\begin{verbatim}
R>  plot(mod.gast, type.plot = "Hazard", plot.mediation= "All")
\end{verbatim}
\end{small}
\begin{center}\includegraphics[width=0.6\linewidth]{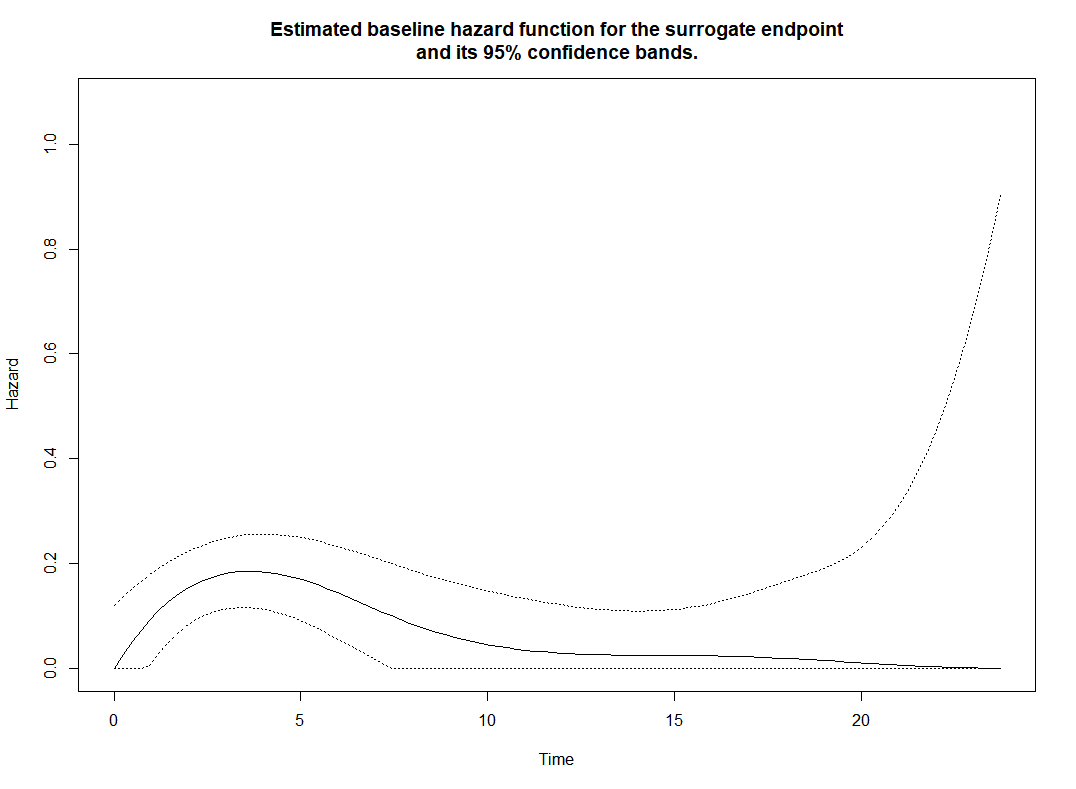} \end{center}
\begin{center}\includegraphics[width=0.6\linewidth]{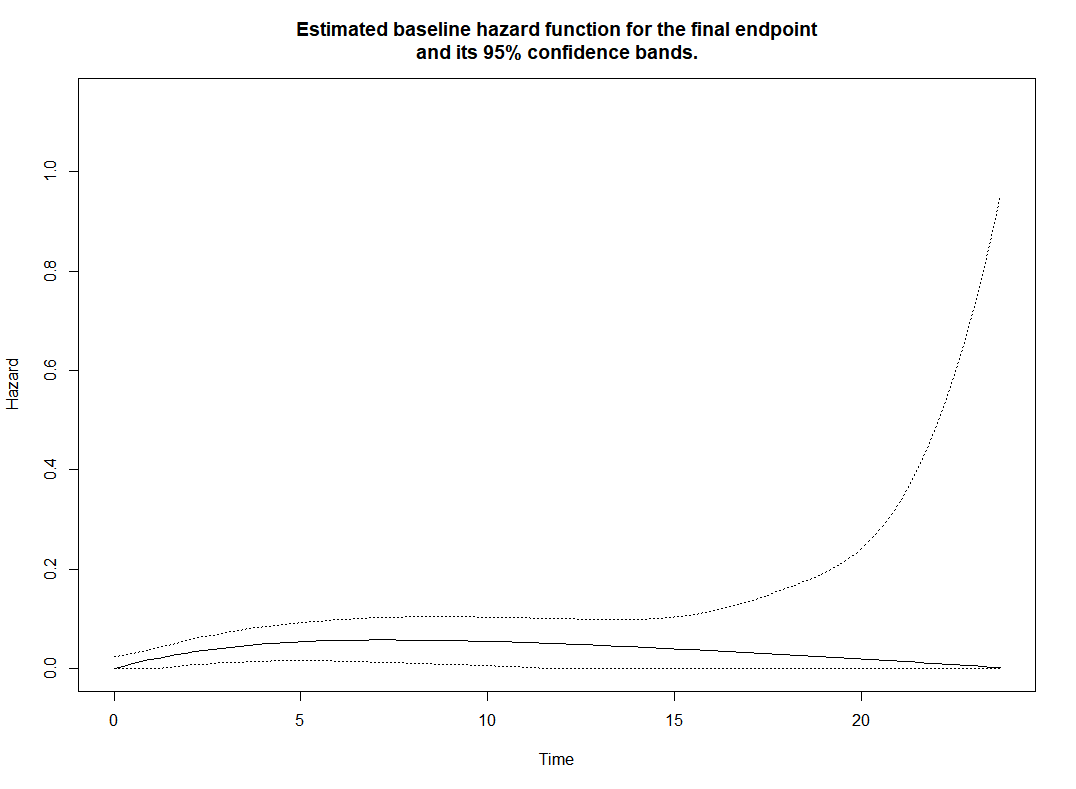} \end{center}
\begin{center}\includegraphics[width=0.6\linewidth]{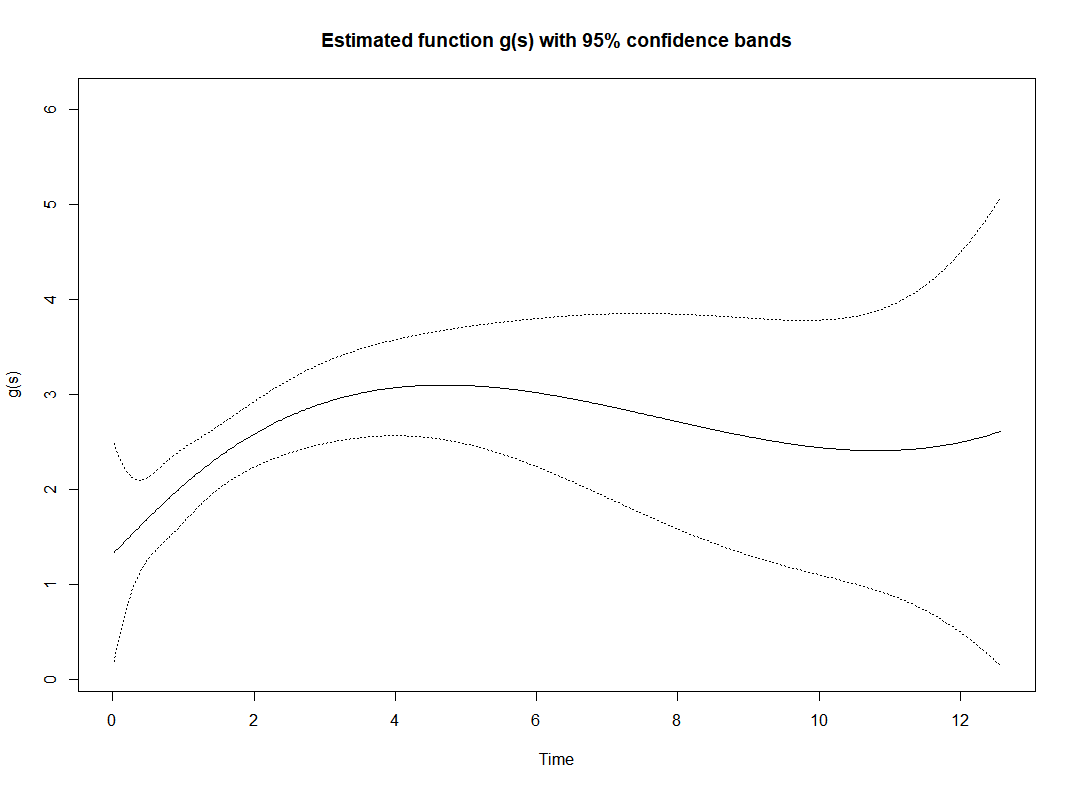} \end{center}
\begin{center}\includegraphics[width=0.6\linewidth]{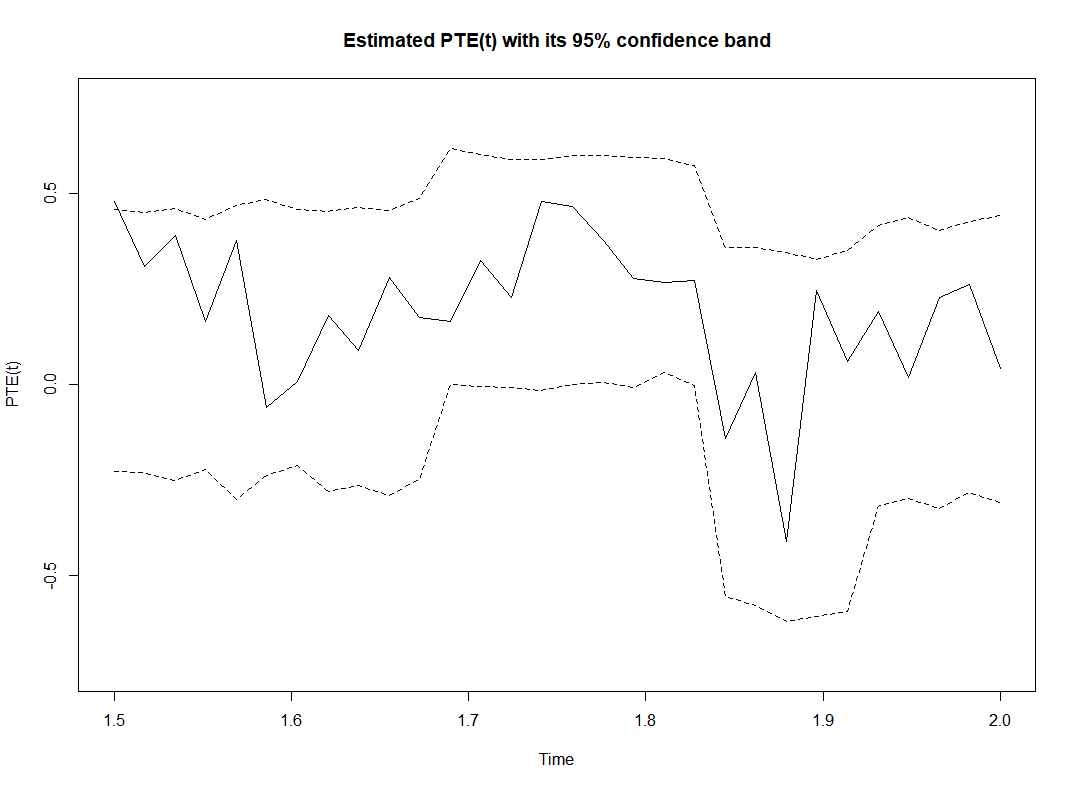} \end{center}
\begin{center}\includegraphics[width=0.6\linewidth]{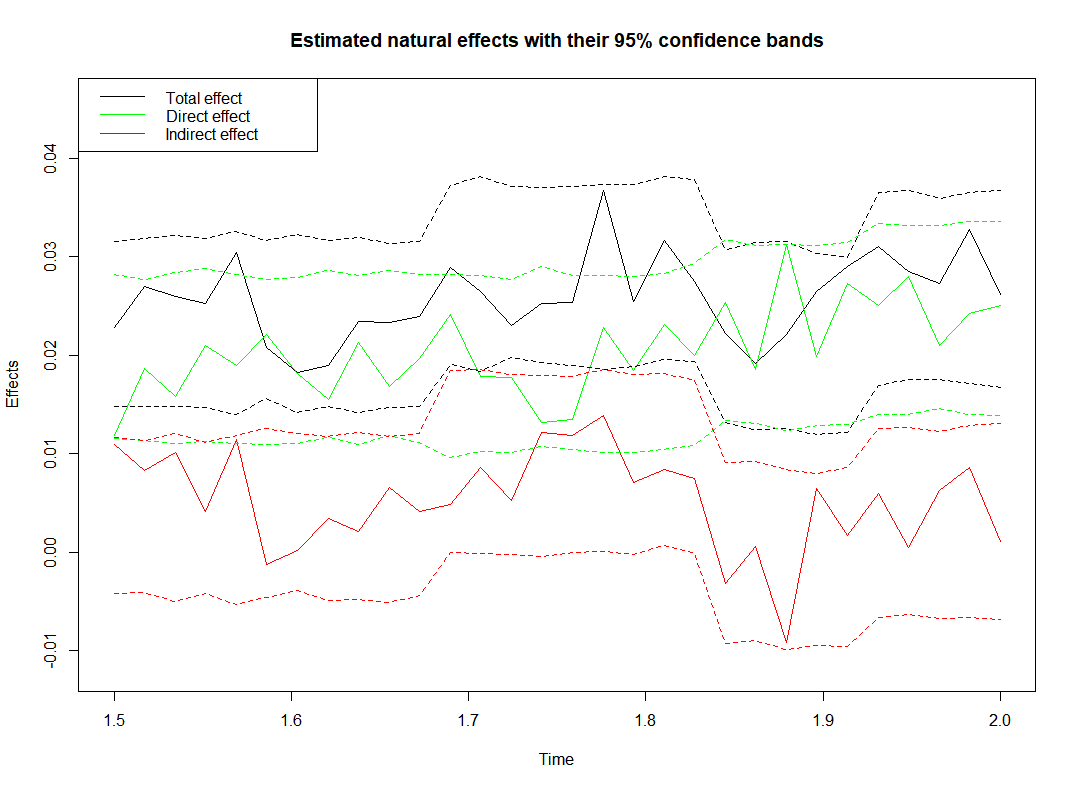} \end{center}

We first have the estimated baseline survival functions associated with
the baseline hazard function \(\hat \lambda_{0,S}(t)\) and
\(\hat \lambda_{0,T}(t)\) and their 95\% confidence bands through the
command \verb|type.plot = "Survival"|. Another option is to set
\verb|type.plot = "Hazard"| to plot the \(\hat \lambda_{0,S}(t)\) and
\(\hat \lambda_{0,T}(t)\). The argument \verb|plot.mediation| can
takes the values \verb|"g"|, \verb|"effects"|, \verb|"pte"| or
\verb|"All"| depending if one want to plot only \(\gamma(\cdot)\), the
natural effects (\(\nie(t)\),\(\nde(t)\) and \(\tte(t)\)), the
\(\pte(t)\) or all at the same time.

From these figures we see that the direct effect is indeed not
significant, while the indirect effect is. However, even if the direct
effect is not significantly different from \(0\), most of its confidence
bands contains negative values therefore the \(\nie\) and \(\nde\) are
in opposite direction. Therefore we have
\(0 \leq \operatorname{TE} \leq \operatorname{NIE}\) and thus
\(\pte \geq 1\). Moreover, the large confidence interval of \(\pte\) can
be explained by the confidence bands of \(\tte(t)\) containing values
very close to \(0\), hence a large instability of \(\pte(t)\).

Finally, there is not enough evidence, for both the association and mediation approaches to suggest good surrogacy in this context. A limitation is the small total treatment effect which makes difficult the interpretation of the
\(\pte(t)\) since it can be very unstable in that case.

\subsection{Tumor size as a surrogate biomarker of overall survival in colorectal cancer: a mediation approach}

In this third application we are interested in evaluating the tumor size
evolution over time as a surrogate of the overall surival in colorectal
cancer. Since the tumor size evolution is a longitudinal biomarker we
will base the analysis on the function \verb|longiPenal|.

We will use a dataset containing 150 patients randomly selected from the
FFCD 2000-05 multicenter phase III clinical trial
\citep{ducreux2011sequential}. This trial originally included 410
patients with metastatic colorectal cancer randomized into two treatment
strategies: combination and sequential chemotherapy. The dataset
contains times of observed appearances of new lesions censored by a
terminal event (death) with some baseline characteristics.

Because the available dataset does not contain the identificator
of the center of the patients and for computational purposes, we illustrate the approach without taking into account the multi-centric nature of the data. 
The data are actually composed of two datasets, one for the survival part
and another containing the repeated measurements of tumor sizes. 

\subsubsection{Dataset}
As for the two previous illustrations these two datasets can be loaded from
\verb|frailtypack|.

\begin{small}
\begin{verbatim}
R>  data(colorectal)
R>  data(colorectalLongi, package = "frailtypack")
\end{verbatim}
\end{small}
The dataset \verb|colorectal| contains several observations per
subject, one for each new lesions in addition to a follow-up time and a
censoring indicator for death.

Therefore we only want to retrieve the last observation for a subject.
In this dataset the variable \verb|new.lesions| takes the value \(1\)
if a new lesion is record and \(0\) otherwise. Therefore if a subject
has \(n_i\) observations, the observations \(1,\dots,n_{i-1}\) all have
the status \verb|new.lesions| equals to \(1\) (since the repeated
follow-up are based on the appearance of new lesions). Hence, the last
observation for each subject can be taken as the only one for which
\verb|new.lesions| equals \(0\):

\begin{small}
\begin{verbatim}
R>  colorectalSurv <- subset(colorectal, new.lesions == 0)
\end{verbatim}
\end{small}
In the dataset the variable \verb|treatment| takes the value
\verb|"S"| for ``sequential'' and \verb|"C"| for ``combined'', for
interpretability we simply make this variable binary 0/1,

\begin{small}
\begin{verbatim}
R>  colorectalSurv$treatment <- sapply(colorectalSurv$treatment,
+   function(t) ifelse(t == "S", 1, 0))
R>  colorectalLongi$treatment <- sapply(colorectalLongi$treatment,
+   function(t) ifelse(t == "S", 1, 0))
\end{verbatim}
\end{small}
To keep the illustration simple we only adjust on the variable
\verb|age| as a categorical variable: \textless60 years, 60-69 years
or \textgreater69 years. 

\subsubsection{Model fitting and surrogacy evaluation}

The call to the function is:

\begin{small}
\begin{verbatim}
R>  mod.col = longiPenal(Surv(time1, state) ~ age + treatment,
+             tumor.size ~ age + year*treatment,
+             data = colorectalSurv, data.Longi = colorectalLongi,
+             random = c("1", "year"), id = "id",
+             link = "Current-level", timevar = "year",
+             method.GH = "Pseudo-adaptive",
+             mediation = TRUE,
+             med.trt = colorectalSurv$treatment,
+             med.center = NULL,
+             n.knots = 7, kappa = 2,
+             pte.times = seq(1,2,length.out = 30),
+             pte.boot = TRUE, pte.nboot = 2000,
+             pte.nmc = 1000)
\end{verbatim}
\end{small}
In this call we fit a model using a ``Current-level'' link function
between the longitudinal biomarker and the final endpoint. We specify a
random slope and intercept in the longitudinal submodel. The arguments \verb|n.knots| and \verb|kappa| specify the number of knots and the penalization term related to the splines baseline hazard function> 

The argument \verb|mediation = TRUE| indicates that we want to compute
the natural direct and indirect effects as well as the proportion of
treatment effect, \(\pte(t)\). We require that this function to be
evaluated at $30$ timepoints between 1 and 2 through the
argument \verb|pte.times|. 
Moreover, we also require that the bootstrap standard error and confidence interval for \(\pte(t)\) computed using \(2000\) samples. Finally, \verb|pte.nmc| specify the number of Monte Carlo sample to be used for integrating over the random effects distributions for the computation of the mediation-related quantities such as the \(\pte(t)\) and the natural direct and indirect effects.

The result can be displayed by applying the \verb|R| function
\verb|print| to the object \verb|mod.col|.

\begin{small}
\begin{verbatim}
R>  print(mod.col)
\end{verbatim}
\end{small}
\begin{small}
\begin{verbatim}
Call:
longiPenal(formula = Surv(time1, state) ~ age + treatment, 
    formula.LongitudinalData = tumor.size ~ 
    age + year * treatment, data = colorectalSurv, 
    data.Longi = colorectalLongi, random = c("1", "year"), 
    id = "id", link = "Current-level", 
    timevar = "year", n.knots = 7, kappa = 2, 
    mediation = TRUE, med.center = NULL, 
    med.trt = colorectalSurv$treatment, 
    method.GH = "Pseudo-adaptive", 
    pte.times = seq(1, 2, length.out = 30), 
    pte.nmc = 1000, pte.boot = TRUE, 
    pte.nboot = 2000)


  Joint Model for Longitudinal Data and a Terminal Event 
  Parameter estimates using a Penalized Likelihood on the hazard function 
  Proportion of treatment effect estimated using mediation analysis  

Longitudinal outcome:
---------------- 
                    coef SE coef (H) SE coef (HIH)         z          p
Intercept       3.137931    0.162309      0.162293 19.333100     <1e-16
age60-69 years  0.018671    0.161011      0.161005  0.115963 9.0768e-01
age>69 years   -0.199278    0.132941      0.132937 -1.498991 1.3388e-01
year           -0.856672    0.119570      0.119537 -7.164634 7.7993e-13
treatment      -0.014291    0.210141      0.210140 -0.068007 9.4578e-01
year:treatment  0.573562    0.179795      0.179791  3.190090 1.4223e-03

      chisq df global p
age 3.64584  2    0.162

Terminal event:
------------- 
                    coef exp(coef) SE coef (H) SE coef (HIH)         z          p
age60-69 years -0.189319  0.827522    0.238821      0.237460 -0.792725 4.2794e-01
age>69 years    0.043271  1.044221    0.221663      0.219158  0.195213 8.4523e-01
treatment      -0.073594  0.929049    0.199693      0.199502 -0.368536 7.1247e-01

       chisq df global p
age 0.984705  2    0.611

 
Components of Random-effects covariance matrix B1: 
                             
Intercept  2.047206 -0.486161
year      -0.486161  0.733352

Association parameters: 
                  coef        SE       z          p
Current level 0.340028 0.0772822 4.39982 1.0834e-05

Residual standard error:  0.928261  (SE (H):  0.025609 ) 
 
Mediation analysis:
------------- 

  
Estimated PTE, natural direct, indirect and total effect at 5 time points 
 
    Time    PTE          CI.95.PTE      TE         CI.95.TE    NDE        CI.95.NDE
1 1.1724 0.4808  [-9.9605;11.7879] -0.0218 [-0.0865;0.0738] 0.0119 [-0.0666;0.1168]
2 1.3793 0.4861 [-13.6501;11.2626] -0.0120  [-0.0865;0.074] 0.0158 [-0.0654;0.1165]
3 1.5862 0.4959 [-12.3904;16.3048] -0.0182 [-0.0868;0.0757] 0.0138 [-0.0652;0.1156]
4 1.7931 0.5212  [-9.9141;14.5458] -0.0201 [-0.0858;0.0739] 0.0167 [-0.0637;0.1186]
5 2.0000 0.8399 [-19.2805;14.5575] -0.0101 [-0.0376;0.0269] 0.0123 [-0.0276;0.0653]
      NIE         CI.95.NIE
1 -0.0337 [-0.0587;-0.0113]
2 -0.0278 [-0.0579;-0.0115]
3 -0.0321 [-0.0572;-0.0119]
4 -0.0367 [-0.0591;-0.0105]
5 -0.0224 [-0.0445;-0.0072]

      penalized marginal log-likelihood = -1668.92
      Convergence criteria: 
      parameters = 1.37e-05 likelihood = 0.000256 gradient = 1.34e-06 

      LCV = the approximate likelihood cross-validation criterion
            in the semi parametrical case     = 1.59769 

      n= 150
      n repeated measurements= 906
      n events= 121
      number of iterations:  14 

      Exact number of knots used:  7 
      Value of the smoothing parameter:  2
      Gaussian quadrature method:  Pseudo-adaptive with 9 nodes
\end{verbatim}
\end{small}
From these results we see a significant effect of the treatment-time interaction on the surrogate (through the slope), while the estimated \(\beta_{Z,T}\) is not significantly different from \(0\). The
individual random effects suggest a heterogeneity at baseline of the
tumor size value. Moreover the estimated covariance matrix suggest no
treatment heterogeneity across the trial, which is expected due to
randomization (of the trials). Moreover the association between the surrogate and the final endpoint (through the \verb|Current-level| parameter) is significantly different from $0$. 

The results regarding the mediation analysis are displayed in the \verb|Mediation analysis| section of the printed results. It shows the estimated \(PTE\), total effect (TE), natural direct effect (NDE) and natural indirect effect (NIE) at several times. While in the function call we specified these quantities to be computed at $30$ time points, the print call returns by default only $5$ of them for readability. From these result we see that \(PTE(t)\) ranges from $0.480$ for $t$ close to $1$ to $0.840$ for $t=2$. However, the estimated total effect (TE) is very close to $0$ at each time and its confidence interval contains $0$ at each time. Therefore, the confidence intervals for \(PTE(t)\) is very broad and no meaningful interpretation of surrogacy can be made in this case.

The estimated baseline hazard function \(\hat \lambda_{0,T}(t)\), \(PTE(t)\) and estimated natural effects can be plotted using the \verb|R| function \verb|plot|.

\begin{small}
\begin{verbatim}
R>  plot(mod.col, plot.mediation = "All", conf.bands = TRUE)
\end{verbatim}
\end{small}
\begin{center}\includegraphics[width=0.6\linewidth]{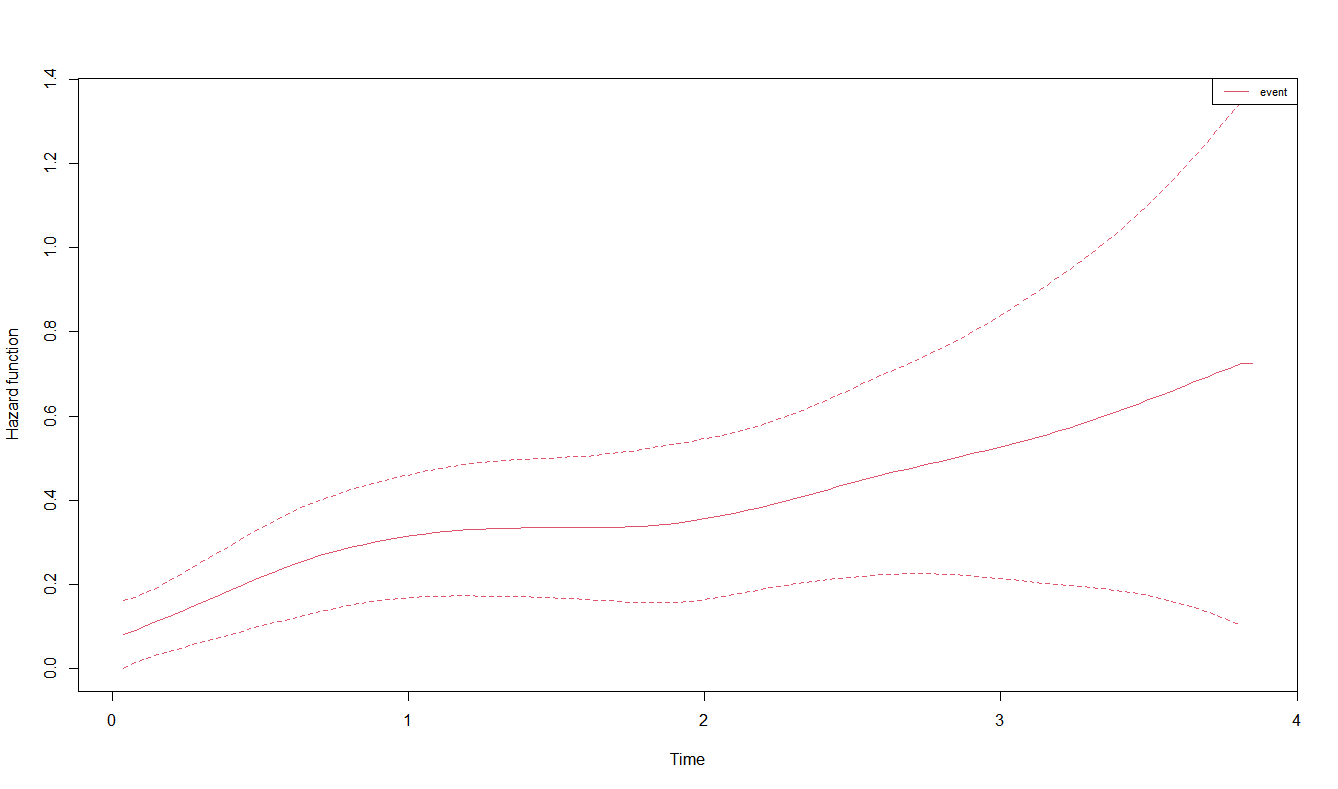} \end{center}
\begin{center}\includegraphics[width=0.6\linewidth]{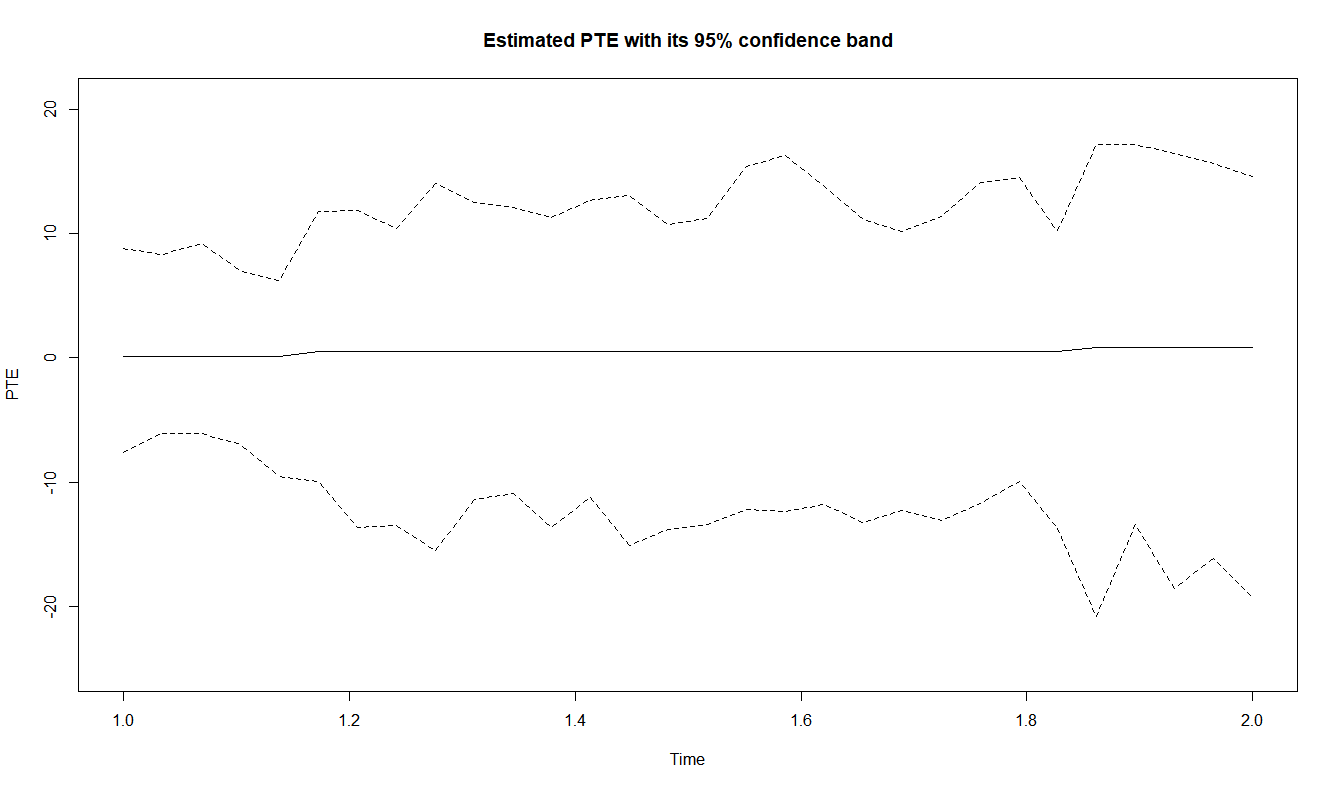}  \end{center}
\begin{center}\includegraphics[width=0.6\linewidth]{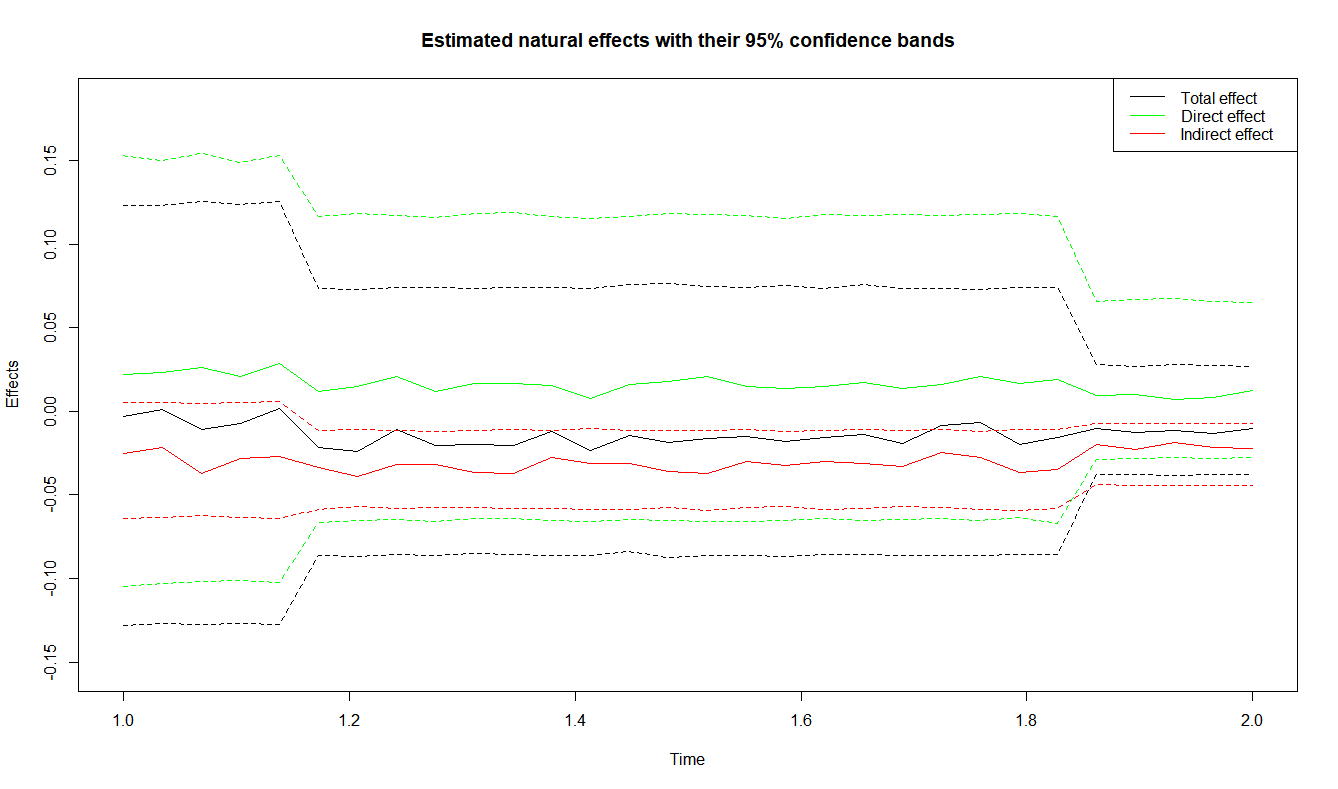}  \end{center}

As previously stated, regarding the mediation measures, we can see that neither the direct effect, \(\nde(t)\), indirect effect \(\nie(t)\) nor the total effect,
\(\tte(t)\) are significant, which results in a very large confidence
interval for the \(\pte(t)\).

As we can see, the confidence bands for the estimated \(\pte(t)\) are
very wide. This is explained by the absence of
treatment effect on the final endpoint; therefore \(\tte(t) \approx 0\)
which results in very high uncertainty regarding \(\pte(t)\). As such,
it is impossible to conclude regarding the surrogacy, 
since the absence of treatment effect on the final endpoint 
renders the search for surrogate endpoints pointless. 
The validation of surrogates endpoint should only be 
made for treatments that have proved their efficacy regarding 
the final endpoint. In this application the low number of patients (150) is a clear
limitation.

\section{Discussion}

In this paper we presented new implementations of surrogate endpoint
validation based on joint models that combines meta-analytic data and
mediation analysis. These joint models have been implemented in the
\verb|R| package \verb|frailtypack|. These models combine the two
main approaches that have been developed in the statistical literature
for surrogate endpoint validation: the meta-analytic approach and the
causal mediation analysis approach.

To achieve a shorter computation time, the core of these functions were
developed in \verb|Fortran| and support multithreading through
\verb|OpenMP|. This parallelization takes place at the numerical
approximation of the integral over the random-effects where the
iterations (either in a Monte Carlo or quadrature approach) are
dispatched between the available threads.

Numerical issues can happen with the estimation of these complex models,
especially when computing the likelihood which requires computing
integrals over the random effects (both individual and trial level).
These integrals do not have closed-form solutions and therefore need to
be numerically approximated. Two main approaches for carrying out these
approximations can be used, either by using Monte Carlo sampling or
Gauss-Hermite quadrature rule. Based on simulations results, it appears
that better numerical approximation and convergence of the maximization
algorithm is achieved when a Monte-Carlo approach is used to integrate
over the trial-level random effects and a Gauss-Hermite rule is used for
the integral over the individual-level random effects. The latter can be
further improved by using pseudo-adaptive quadrature in which the nodes
in the quadrature are chosen to properly recover the scale and shape of
the integrand \citep{rizopoulos2012fast}.

Another computational issue concerns specifically Model
(\ref{eq:model2}). The likelihood of this model requires computing the
product of the individual-likelihoods \(\prod_{j=1}^{n_i} L_{ij}\) for
all the patients in trial \(i\). Numerical instabilities can occur if
\(n_i\) is large and/or if the number of observations per subject is
large. If all individual likelihoods are small, say
\(L_{ij} \approx 10^{-3}\), and if \(n_i\) is large, \(n_i \approx 200\)
as can be the case for clinical trials, then the product will result in
a numerical underflow and rounded to \(0\). One possible way around is
to use a large constant, \(M\), and multiply each individual likelihood
by \(M\) so that there is no longer a numerical underflow. Since \(M\)
is a constant it can then be easily taken out of the integral and
substracted (when taking the log of this integral). However, in an
iterative maximization algorithm, \(M\) may be suited for the first
iterations but as the vector of parameter is updated, so will be the
``scale'' of the individual-likelihoods \(L_{ij}\) which depend on the
parameters. Therefore \(M L_{ij}\) may be too large at a given iteration
which can also result in numerical issues (numerical overflow this
time). One possibility is to also update to value of \(M\) at each
iteration so that it keeps track of the scale of the individual
likelihoods.

Regarding the individual surrogacy measure (the Kendall's \(\tau\)), the
introduction of a direct effect of the surrogate on the final endpoint
creates a difficulty regarding both its definition and computation since
this link influences the association between the two endpoints and
therefore must be taken into account. For example, in Model
(\ref{eq:model1}), the Kendall's without a direct link between \(S\) and
\(T\) (i.e.~without the function \(\gamma(S)\)), is given by Equation
(\ref{eq:tau2}) which only requires integrating over the distribution of
the random effects while the more general \(\tau\) in the presence of
\(\gamma(S)\) is given by Equation (\ref{eq:tau1}) and requires more
complex integration over both the distribution of \(S\) and the
distribution of the random effects. For the trial-level association
measure, \(R^2_{\text{trial}}\), this issue does not occur since the
trial-level random effects take into account all the association between
the treatment effects on the surrogate and the final endpoint.

For now the \verb|jointSurroPenal| function does not allow adjustment on
covariates beside the treatment itself, and the surrogacy measures
(especially the Kendall's \(\tau\) and the \(\pte(t)\)) are defined and
computed in a somewhat ``covariate-free'' manner. Further developments
will allow the adjustment on potential covariates in the model and the
computation of these surrogacy measures conditionally to a given set of
covariates as well as at a marginal (populational) level
\citep{emura2021conditional}. Adjustment on covariates is already
possible in the \verb|longiPenal| function. However it should be noted
that computing a marginal measure averaged over the distribution of the
covariates can significantly increase the computation time in the
presence of continuous variables. Indeed, marginalization in the case of
categorical covariates only requires to compute a \(\tau\) or
\(\pte(t)\) for each possible group and then averaging them (and
weighting by the proportion of each group in the dataset). On the other
hand, marginalization in the presence of a continuous covariates would
require computing these measures for each observed values of the
covariate in the dataset (i.e.~for each individual) which can
significantly increase the computation time.

Unlike the meta-analytic approach, the mediation analysis does not
require, \emph{per se}, the use of meta-analytic data. The combination
of meta-analytic data and mediation analysis allows broader
generalisation of the results by introducing heterogeneity in the data.
If one only has access to single-trial data, a possibility is to use
different clustering units instead of ``trials'', such as
``institution'' in multicentric studies. Even if our proposed approaches
were developed within a meta-analytic setting which is more appropriate
for surrogate validation, a further development would be to run a
mediation analysis without the requirement of meta-analytic or clustered
data (which would yield a simpler model). However, in that case only the
individual-level association in addition to the proportion of treatment
effect would be available since the trial-level association would not be
identifiable.

We want to point out that in a perspective of surrogate validation,
there is no acknowledged threshold yet for the surrogacy measures,
especially the proportion of treatment effect, above which a surrogate
would be claimed validated. Such threshold warrants more investigation
in order to give a clear decision rule for the clinicians and
statisticians.

Finally, further developments of the \verb|frailtypack| package will
concern the extension of the proposed functions to validate surrogate
endpoints. In order to improve the flexibility of the proposed
approaches, other options for the type of surrogate or final endpoint
will be proposed, for example in the case of a binary final endpoint.

\printbibliography

\section*{Acknowledgements} 
Q. Le Coënt and C. Legrand acknowledge the support of the ARC project IMAL (grant 20/25-107) financed by the Wallonia-Brussels Federation and granted by the Académie universitaire Louvain.

\end{document}